\documentclass[journal,final,twocolumn,10pt,twoside]{IEEEtran}
\usepackage{lineno,hyperref}
\usepackage{graphicx}
\usepackage{amsmath,amssymb,amsthm}
\usepackage{booktabs}
\usepackage{balance}
\usepackage{wrapfig}
\usepackage{lipsum}
\usepackage{xcolor}
\usepackage{siunitx}
\graphicspath{{images/}{../images/}}

\DeclareMathOperator*{\argmax}{arg\,max}

\usepackage{subfiles}

\begin{document}









\title{Towards High Data-Rate Diffusive Molecular Communications: Performance Enhancement Strategies}


\author{Mustafa~Can~Gursoy$^\dagger$, Masoumeh~Nasiri-Kenari$^\ddagger$, and Urbashi~Mitra$^\dagger$\\
$^\dagger$ University of Southern California, $^\ddagger$ Sharif University of Technology
}

\markboth{Submitted Paper}
{Submitted Paper}

\maketitle

\begin{abstract}
Diffusive molecular communications (DiMC) have recently gained attention as a candidate for nano- to micro- and macro-scale communications due to its simplicity and energy efficiency. As signal propagation is solely enabled by Brownian motion mechanics, DiMC faces severe inter-symbol interference (ISI), which limits reliable and high data-rate communications. Herein, recent literature on DiMC performance enhancement strategies is surveyed; key research directions are identified. Signaling design and associated design constraints are presented. Classical and novel transceiver designs are reviewed with an emphasis on methods for ISI mitigation and performance-complexity tradeoffs. Key parameter estimation strategies such as synchronization and channel estimation are considered in conjunction with asynchronous and timing error robust receiver methods. Finally, source and channel coding in the context of DiMC is presented.
\end{abstract}

\section{Introduction}
\label{sec:introduction}

\par Molecular communication (MC) is a bio-inspired communication approach that conveys information using chemical signals \cite{firstpaper_lookslike}. In an MC link, signal propagation can be realized through various biochemical mechanisms, including diffusion, active transport, bacteria, calcium signaling, \textit{etc.} \cite{akyildiz_nanocomnet,molecularbook,survey_farsad_2016}. Among these methods, diffusive molecular communications (DiMC) has gained particular interest due to its energy efficiency and bio-compatibility.

\par In a DiMC system, the molecules rely solely on diffusion dynamics after their release from the transmitter. Each emitted molecule exhibits Brownian motion in the channel \cite{molecularbook}, which causes its arrival time at the receiver to be stochastic \cite{survey_schober_2019}. In a time-slotted DiMC link, this stochastic arrival time may cause some molecules to arrive at the receiver later than their intended interval, causing the well-known inter-symbol interference (ISI) issue of DiMC. 



\par In its current stage of research, diffusive molecular communications provide notoriously low data rates due to ISI. These low rates might be acceptable for some applications where the information to be transmitted is limited and the design goals target low-complexity and low energy consumption. Several examples to these applications are one-shot DiMC or DiMC with very long symbol durations \cite{oneshot1,oneshot2}, anomaly detection \cite{abnormality1,abnormality3}, entity localization \cite{localization,localization2}, bio-sensing applications, \textit{etc.} On the other hand, at both micro- and macro-scales, employing DiMC for digital communication links naturally elicits interest in increasing data rates. In fact, such an enhancement can also expand the reach of DiMC into a wider spectrum of applications. To this end, we discuss performance enhancement strategies for DiMC systems. We define \textit{performance enhancement} as gains in one or more of the following:
\begin{itemize}
    \item Lower error probability at the same data rate and comparable transceiver complexity,
    \item Higher data rate at the same target error probability and comparable transceiver complexity,
    \item Lower transceiver complexity at comparable error performance and data rate,
    \item Increased robustness against channel estimation and synchronization errors.
\end{itemize}
Motivated by these goals, we discuss some of the recent advancements in transceiver design that are proposed to enhance communication performance in DiMC. Specifically, the paper covers
\begin{enumerate}
    \item signaling degrees of freedom for DiMC,
    \item recent information theoretic advancements on DiMC systems,
    \item detection and equalization schemes, 
    \item channel estimation and synchronization schemes, alongside non-coherent and asynchronous detectors,
    \item recent channel coding approaches.
\end{enumerate}
In addition, considering the current state of research endeavors in these topics, we further discuss potential avenues for future research.

\par The rest of the paper is organized as follows: Section \ref{sec:LTI_poisson_section} presents the considered system model and its channel characteristics. Section \ref{sec:degrees_of_freedom} discusses signaling degrees of freedom and their design considerations. Section \ref{sec:channel_capacity} presents several recent studies on the information theoretic treatment of the DiMC channel. Section \ref{sec:transceiver_SP} addresses the DiMC system from a design perspective by discussing various transceiver strategies including schemes for detection, equalization, channel estimation, and synchronization. Section \ref{sec:channel_coding} discusses the recent coding literature and the design considerations for future research in this sub-field. Section \ref{sec:PerformanceEval} provides error performance comparisons of the schemes presented in Section \ref{sec:transceiver_SP}. Lastly, Section \ref{sec:conclusion} talks about future research directions and design considerations going forward, and gives the concluding remarks.


\section{DiMC Channel Characteristics}
\label{sec:LTI_poisson_section}

\par To focus our attention on specific transceiver strategies for molecular diffusive communication, we specify the channel model under consideration.  The starting point of most channel modeling efforts is Fick's laws of diffusion for which, the single particle case describes classical Brownian motion \cite{molecularbook}. A key goal of channel modeling efforts is to develop descriptions that enable analysis and design.  To this end, our goal is a stochastic model to capture the unique challenges of the DiMC. We point the reader to \cite{survey_schober_2019} which provides a comprehensive survey of channel modeling techniques.

\par We focus on a point-to-point DiMC link between a point transmitter and a perfectly absorbing spherical receiver in a three dimensional, unbounded environment. It is assumed that no object other than the transmitter and the receiver exist in the communication environment, and the channel is time-invariant (\textit{i.e.} transmitter and/or receiver are not mobile, temperature changes are negligible). 
\begin{figure}[!t]
	\centering
	\includegraphics[width=0.4\textwidth]{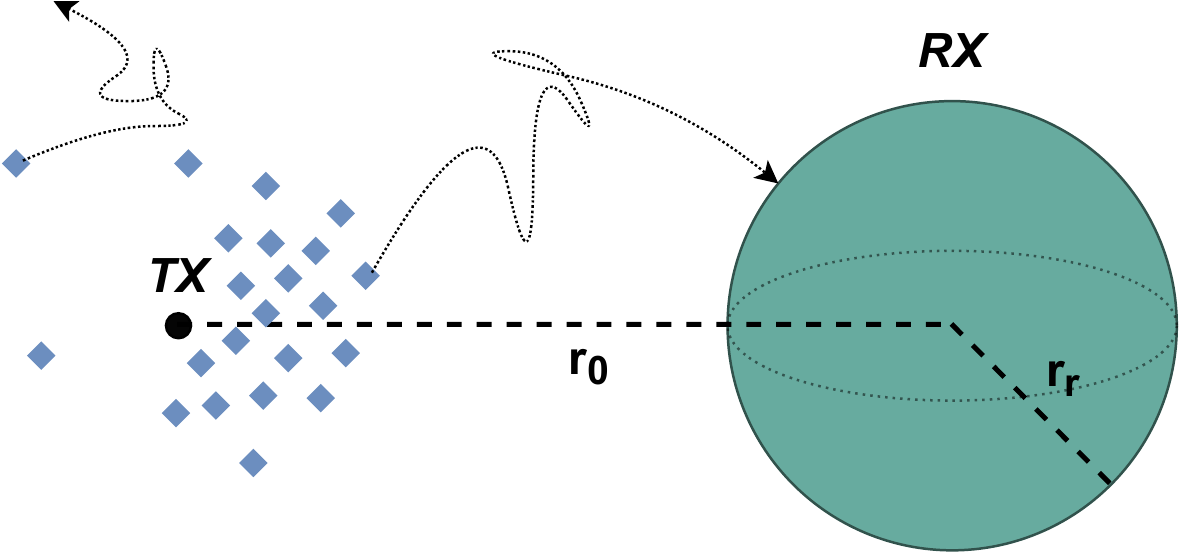} 
	\caption{The system model of interest.}
	\label{fig:systemmodel}
\end{figure}
Denoting the receiver radius as $r_r$, the point-to-center distance between the transmitter and receiver as $r_0$, and the diffusion coefficient of the messenger molecules as $D$, the time arrival density of molecules is given in \cite{3Dchar} as
\begin{equation} \label{eq:arrival_pdf}
	f_{hit}(t) = \frac{r_r}{d+r_r} \frac{1}{\sqrt[]{4\pi Dt}} \frac{d}{t} e^{- \frac{d^2}{4Dt} }, \hspace{1cm} t \in (0,\infty),
\end{equation}
where $d = r_0 - r_r$. Taking the integral of \eqref{eq:arrival_pdf} with respect to time yields the cumulative arrival function, $F_{hit}(t)$, given by
\begin{equation}\label{eq:arrival_cdf}
F_{hit}(t) = \frac{r_{r}}{r_{0}} \text{erfc}\bigg( \frac{r_{0}-r_{r}}{\sqrt[]{4Dt}} \bigg), \hspace{1cm} t \in (0,\infty),
\end{equation}
where $\text{erfc}(\cdot)$ is the complementary error function for a standard Gaussian random variable with mean zero and unit variance. The $F_hit(t)$ function denotes the probability of a single molecule's arrival within the interval $(0,t]$. One interesting observation from \eqref{eq:arrival_cdf} is that 
\begin{equation}
    \label{eq:limitF}
    \lim_{t\rightarrow \infty} F_{hit}(t) = \frac{r_{r}}{r_{0}},
\end{equation}
implying that in an unbounded 3-D environment, there is a non-zero probability ($1 - \frac{r_{r}}{r_{0}}$) that a molecule {\em never} arrives at the receiver. Note that this phenomenon translates to an inherent propagation loss from a communications engineering standpoint.

\par
\begin{figure}[!t]
	\centering 
	\includegraphics[width=0.48\textwidth]{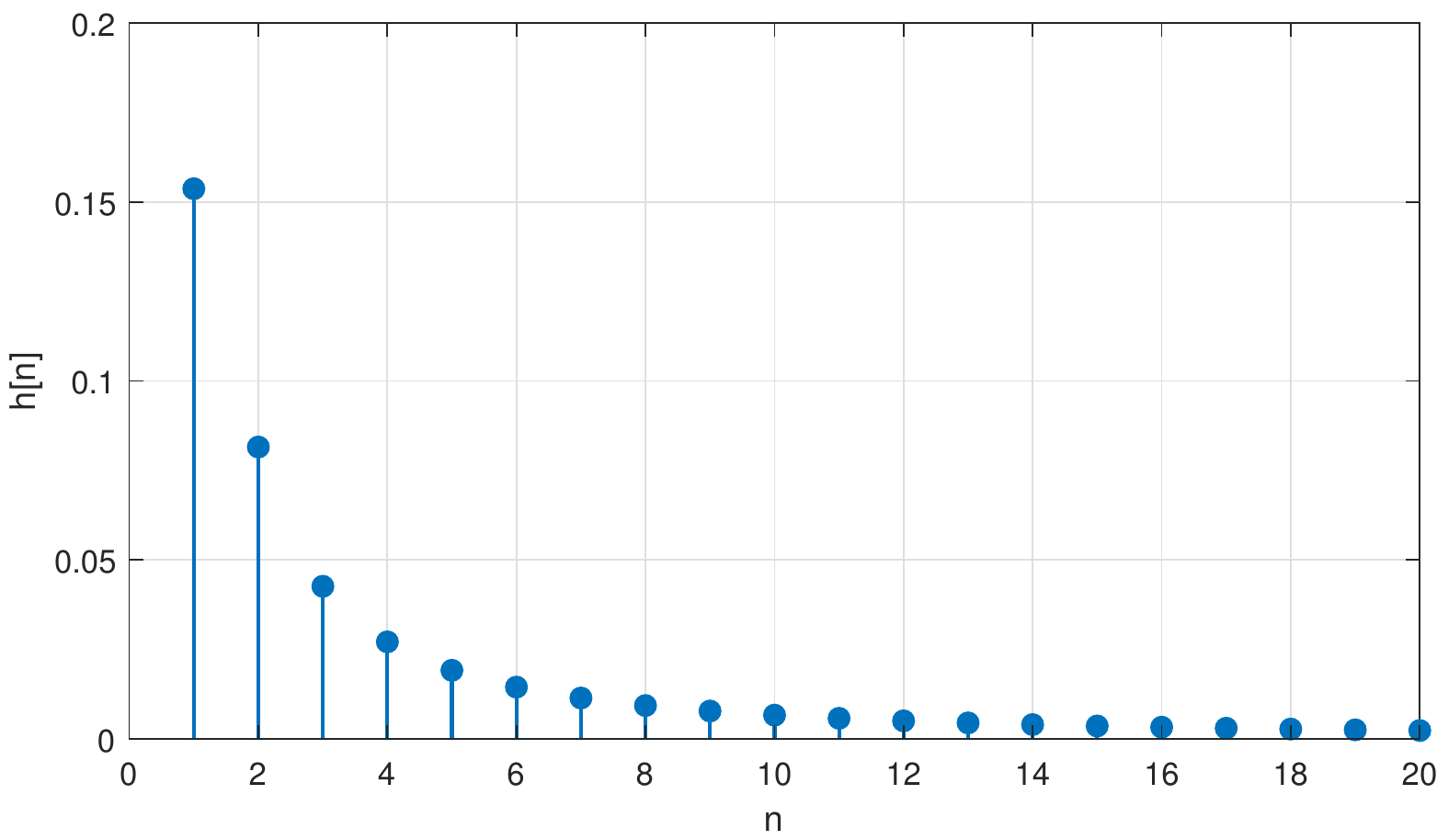} 
	\caption{First $L=20$ channel coefficients of an exemplary DiMC system. $r_0 = \SI{10}{\micro\meter}$, $r_r = \SI{5}{\micro\meter}$, $D = 80 \frac{\SI{}{\micro\meter\squared}}{\SI{}{\second}}$, $t_b = \SI{0.15}{\second}$, and $N=1$.}
	\label{fig:h_vector}
\end{figure}

 We consider a time-slotted digital DiMC, where the time is divided into slots of length $t_s$. This defines a discrete-time channel, which is characterized by the channel coefficients $h[n]$ where $n \in \mathbb{Z}^{+}$. Here, $\mathbb{Z}^{+}$ denotes the set of positive integers. The channel coefficient vector $\boldsymbol{h}$ is obtained from $F_{hit}(t)$ by computing
\begin{equation}\label{eq:FIRcoefs}
h[n] = F_{hit}(n t_s) - F_{hit}((n-1)t_s), \hspace{1cm} n \in \mathbb{Z}^{+}.
\end{equation}
Note that there exists, effectively, an infinite number of channel coefficients, as $f_{hit}(t)$ has a heavy right tail \cite{3Dchar}. However, for modeling purposes, $\boldsymbol{h}$ is considered as a vector of $LN < \infty$ elements, where $L$ denotes the effective \textit{channel memory} in symbols, and $N$ is the number of samples taken by the receiver per one symbol interval. Here, $L$ needs to be sufficiently large to capture the bulk of the heavy right tail of \eqref{eq:arrival_pdf}. Note that the fact that $h[n]$ is non-zero for $n > N$ suggests that the DiMC channel is subject to inter-symbol interference (ISI). An exemplary channel coefficient vector $\boldsymbol{h}$ is presented in Figure \ref{fig:h_vector} to visually illustrate the channel characteristics as a function of time. To provide context, for the scenario considered in Figure \ref{fig:h_vector}, $50\%$ of the molecules never hit the receiver (see Equation \eqref{eq:limitF}) and about $35\%$ of the molecules that do arrive, arrive after the first five time slots, underscoring the heaviness of the right tail.

\begin{figure*}[!t]
	\centering
	\includegraphics[width=0.7\textwidth]{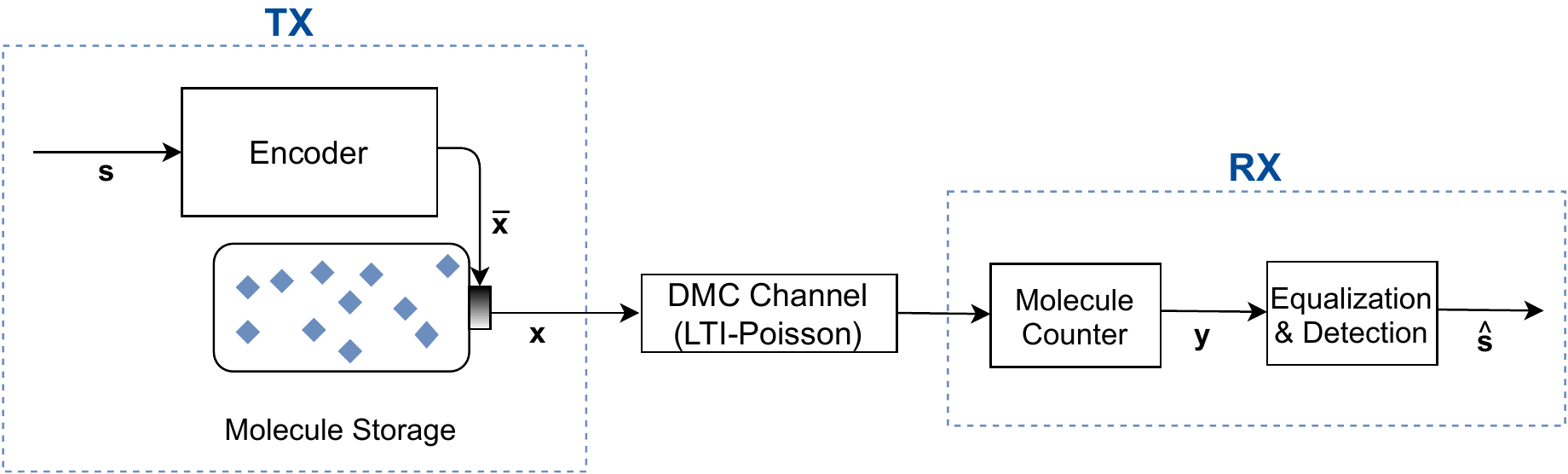} 
	\caption{Overall block diagram of the considered DiMC system.}
	\label{fig:blockdiagram}
\end{figure*}

\par For a non-mobile point-to-point DiMC link in the absence of external noise, the vector  $\boldsymbol{h}$ completely characterizes the channel. In this paper, we also consider an external Poisson noise with rate $\lambda_s$ per each sample. Denoting the vector that holds the number of molecules emitted by the transmitter as $\boldsymbol{x}$, the number of molecules arriving at the $n^{th}$ time slot is approximately distributed as 
\begin{equation} \label{eq:y_seq}
	y[n] \sim \mathcal{P}( \Big( \sum_{k = 1}^{LN} h[k] x[n-k+1] \Big) + \lambda_s ),
\end{equation}
where $\mathcal{P}(\mu)$ denotes the Poisson distribution with mean rate parameter $\mu$. This model is referred to as the Linear-Time-Invariant (LTI)-Poisson model \cite{LTI_Poisson}. The acronym LTI refers to the fact that the rate of the Poisson random variable is the convolution of the hitting probabilities with the transmitted sequence. It is also common to further approximate the Poisson arrivals with Gaussian random variables \cite{arrivalmodel}, where 
\begin{equation}
    \label{eq:y_seq_Gauss}
    \begin{split}
        y[n] &\sim \mathcal{N}(\mu[n],\mu[n]) \\
        \mu[n] &= \Big( \sum_{k = 1}^{LN} h[k] x[n-k+1] \Big) + \lambda_s.
    \end{split}
\end{equation}
We note that in the LTI-Poisson model, conditioned on the input information sequence, the transmission sequence $\boldsymbol{x}$ can also be a random vector, which is motivated by the nano-scale transmitter's imperfect release behavior \cite{LTI_Poisson}. Specifically, $x[k] \sim \mathcal{P}(\bar{x}[k])$, where the emission rate (mean) vector $\bar{\boldsymbol{x}}$ is a deterministic function of the symbol sequence $\boldsymbol{s}$ according to the employed modulation scheme. We shall consider both the LTI-Poisson channel as well as its Gaussian approximation model herein. Overall, the block diagram of the considered DiMC link is presented in Figure \ref{fig:blockdiagram}.

\par The LTI-Poisson model with a perfectly absorbing receiver is an idealized approximation to the actual DiMC channel behavior. An exact end-to-end model would include the imperfections in transmission and reception procedures, as well as the time-varying channel conditions. That said, the LTI-Poisson model provides a tractable approximation of the channel statistics, and captures the key phenomena that govern the error performance of a DiMC system: ISI due to the heavy tail of $f_{hit}(t)$, and the consequences of data-dependent arrival statistics.

\section{Signaling Degrees of Freedom}
\label{sec:degrees_of_freedom}

\par In this section, we discuss several signaling degrees of freedom that can be exploited to convey information in a DiMC system. We observe that the exploitation of different degrees of freedom to maximize information transfer is a classical communications engineering problem \cite{landau1966optimality,slepian1976bandwidth,weber2012elements}.  The modulations and signal designs for diffusive molecular communications inherit much from those for radio frequency communications, but also exploit features that are unique to the DiMC \cite{survey_goldsmith_2020}.

\subsection{Bit Duration and Transmission Power Constraints}
\par When comparing different modulation schemes and error control coding strategies, the information bit rate and transmission power per bit need to be normalized for fair comparison. We refer to the first normalization as the \textit{bit duration constraint}, and characterize the constraint by the parameter $t_b$ (seconds per information bit). The latter normalization is referred to as the \textit{transmission power constraint}, and this constraint is characterized by the parameter $M$ (emitted molecules per information bit).

\par For a scheme that can transmit $B$ bits per symbol, these normalizations imply that the scheme is allowed to transmit at a symbol duration of $t_{sym} = B t_b$ and emit an average of $B M$ molecules per each symbol. Note that due to the arrival statistics characterized in \eqref{eq:arrival_pdf}-\eqref{eq:y_seq}, emitting at a larger $t_{sym}$ while satisfying the bit duration constraint is of particular interest, as the ISI for larger $t_{sym}$ would be less. 

\subsection{Emission Intensity}

\par 
Emission intensity, or {\em concentration}, is the basic and the most fundamental means of molecular signaling in nature \cite{molecularbook} and is reminiscent of amplitude modulation is classical communications \cite{roder1931amplitude}. To date, concentration signaling appears to be the most widespread modulation in the DiMC literature, with a focus on binary signaling, (binary concentration shift keying, BCSK) \cite{CSK_MOSK_akyildiz_2011}. This consideration mainly stems from its simplicity and the fact that higher order CSK schemes are outperformed by BCSK in terms of error performance \cite{MCSK_arjmandi_2013}. Throughout the paper, due to its widespread use and simplicity, we use BCSK as the default modulation scheme. We specifically employ its on-off keying (OOK) variant, and use the terms BCSK and OOK interchangeably. 

\par
\begin{figure*}[!t]
	\centering
	\includegraphics[width=0.99\textwidth]{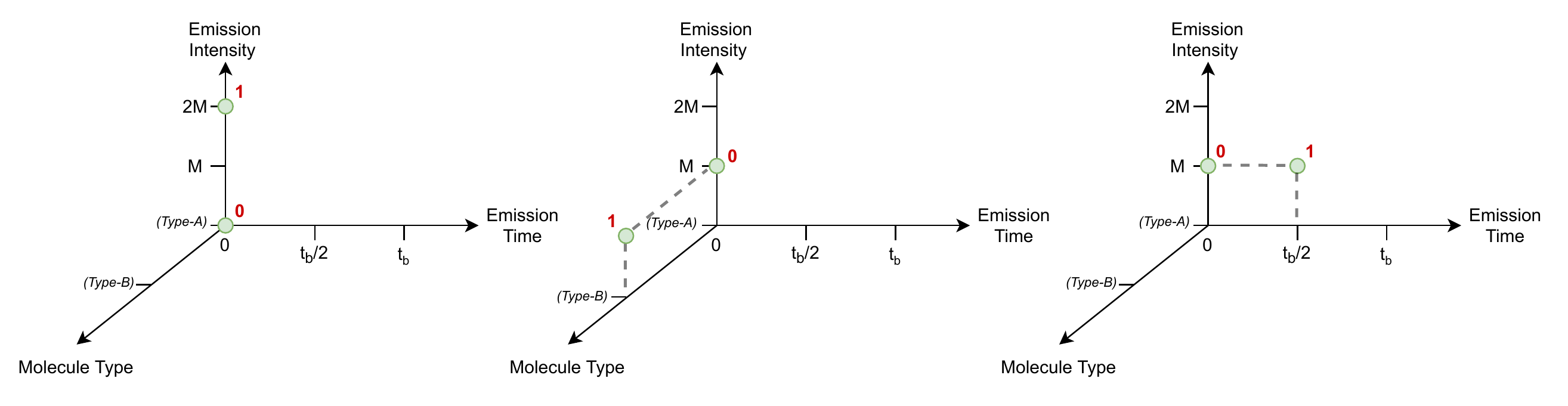} 
	\caption{Discrete-time DiMC constellation diagrams of binary modulations on fundamental degrees of freedom: Emission intensity (BCSK, left), molecule type ($2$-MoSK, middle), emission time ($2$-PPM). Note that binary modulation implies $t_{sym} = t_b$.}
	\label{fig:modulation_diagram}
\end{figure*}

 For the BCSK scheme, the symbol duration $t_{sym} = t_b$ as only one bit is transmitted per symbol. In the paper, we assume equiprobable transmissions of bit-$1$ and bit-$0$. With this assumption, using the OOK variant corresponds to  emitting $2M$ and $0$ molecules for bit-$1$ and bit-$0$, respectively, as illustrated in Figure \ref{fig:modulation_diagram}. That said, we will discuss in Subsection \ref{subsec:detector} that the number of emitted molecules for bit-$1$ can be adjusted in order to mitigate ISI \cite{FTDadaptivetransmission_mitra_2016,ISI_burcu_2015}. Note that even though the scheme in \cite{FTDadaptivetransmission_mitra_2016,ISI_burcu_2015} conserves the binary intensity modulation structure, it is implemented via a finite state machine which exploits channel knowledge at the transmitter side.


\subsection{Molecule Type as a Degree of Freedom}
\label{subsec:type_modulations}

\par Using multiple types of molecules naturally equips the DiMC system with more degrees of freedom over single-molecule strategies. This ability to employ different molecules is a unique aspect of DiMC in contrast to radio frequency based communications. Of course, the price to pay is the increased device complexity required for storing, possibly synthesizing, and sensing different types of molecules. Therefore, the question of how to best utilize this additional degree of freedom arises. One way of utilizing the additional molecule type(s) for performance enhancement is to design more sophisticated modulation schemes. Assuming different molecule types interact negligibly in the channel, each additional molecule type can be thought of operating in an additional,  {\em orthogonal} channel\footnote{We note that in addition to having negligible collisions, different types of molecules should also not react with each other to assume orthogonality. We will note later in this subsection that chemical reactions can be deliberately included in design and this orthogonality can be broken, in order to enhance performance.}. Therefore, the most natural extension of any DiMC modulation scheme to multiple molecules is to send parallel streams of said modulation with each molecule type. In this context, the natural extension of BCSK is the so-called depleted MoSK (D-MoSK) scheme \cite{DMoSK_2015}. Assuming the system has access to two types of molecules, D-MoSK can double the symbol duration while keeping the same bit rate, effectively combating the ISI problem. Note that the receiver counts for the whole $2t_b$ duration, \textit{i.e.} until the next transmission with same molecule. Note that one might deliberately limit oneself to only considering the first $t_b$ duration after release for molecule counting, which corresponds to the molecular concentration shift keying (MCSK) \cite{MCSK_arjmandi_2013}. 

\par In essence, the goal of parallel streams is to \textit{avoid} ISI, given the different molecule types at hand. This concept is characterized in \cite{ISIavoiding_MNK}, where given the channel memory length $L$, the transmitter emits the same molecule type at times slots at least $L$ symbol durations apart. Note that this strategy imposes a constraint on the channel input sequence, which the authors characterize. One can interpret the approach in \cite{ISIavoiding_MNK} as a source coding method by specifying permissible patterns at the channel input. We consider this feature again, in more detail in Section \ref{sec:channel_coding}.


\par In addition to simply repeating the same modulation in parallel channels, multiple molecule types also allow for more sophisticated modulation designs. Assuming $K = 2^B$ for some $B \in \mathbb{Z}_{+}$, each molecule type can be thought of representing an $B$-bit symbol. 
\begin{figure}[t]
	\centering
	\includegraphics[width=0.3\textwidth]{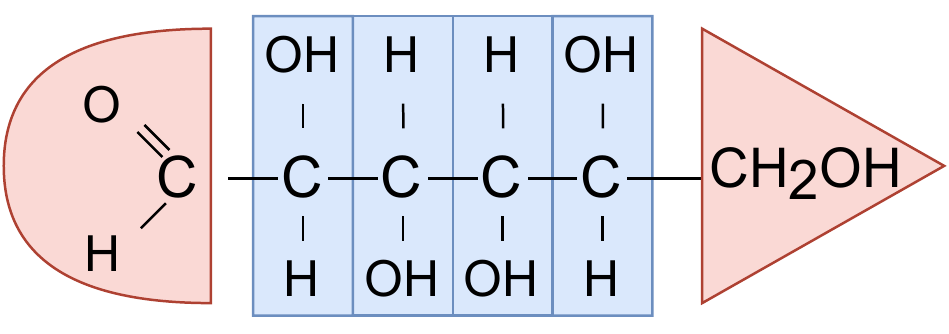} 
	\caption{A D-Galactose molecule can hold $B=4$ bits, ``1001" \cite{isomerMoSK,MaaF_gursoy_2018}.}
	\label{fig:MaaFmolecule}
\end{figure}
This leads to the well-known type modulation of DiMC: the $K$-ary molecular shift keying ($K$-MoSK, \cite{CSK_MOSK_akyildiz_2011}). Standard molecule shift keying is generalized in \cite{GMoSK_2020}, where $K_A$ out of $K$ molecules are simultaneously activated at each transmission. Note that this generalized MoSK (GMoSK) is able to encode
\begin{equation}
    \label{eq:gmosk}
    B = \Big\lfloor \log_2 \binom{K}{K_A}  \Big\rfloor,
\end{equation}
bits in a single symbol, corresponding to having $t_{sym} = \lfloor \log_2 \binom{K}{K_A} \rfloor  t_b $, which greatly reduces ISI. Furthermore, molecular transition shift keying (MTSK, \cite{ISI_burcu_2015}) employs a one-bit memory at the transmitter side and sends BCSK pulses with different molecule types depending on the next bit. This way, MTSK can utilize the constructive interference of consecutive bit-$1$ transmissions, while mitigating ISI for a bit-$0$. Furthermore, \cite{runlength_kwak_2014} considers two types of molecules where one molecule type determines the concentration symbol, whilst the other one encodes the run-length of the said symbol.

\par A molecule family of $K$ distinguishable molecules defines a $\lfloor\log_2(K)\rfloor$-bit long data frame. This frame can be considered to have $b_I$ information bits, and $b_F$ bits that can be used to generate $2^{b_F}$ parallel transmissions of ${b_I}$-bit symbols (\textit{i.e.} $2^{b_I}$-MoSK), where 
\begin{equation}
    \lfloor \log_2(K) \rfloor = b_I + b_F.
\end{equation}
This consideration leads to the {\em molecule-as-a-frame} (MaaF) strategy, where \cite{MaaF_gursoy_2018} shows the existence of an optimal ($b_F$,$b_I$) allocation that minimizes the BER. Note that similar to GMoSK \cite{GMoSK_2020}, MaaF is another generalization of the standard MoSK scheme, where $\lfloor \log_2(K) \rfloor = b_I$ corresponds to $K$-MoSK. It is shown in \cite{GMoSK_2020} that MaaF outperforms GMoSK in the low signal-to-noise ratio (SNR) regime, whereas GMoSK is the desirable scheme for high signal-to-noise ratio (SNR).

\par Additional molecule types can also be used for equalization purposes. For example, in a dual-molecule DiMC system, upon emission of a type-A molecule, one can select type-B as the enzyme that degrades type-A \cite{enzymes_noel_2014}. This way, type-B molecules degrade the type-A molecules that linger in the channel for longer times, effectively mitigating ISI. Furthermore, as shown in \cite{twowayrelay_MNK}, degradation reactions can also be employed at various stages of a two-way molecular communications link to mitigate ISI in both directions. 

\par The approach in \cite{aminusb_burcu_2015} shows that a degradation-like effect can be achieved even without molecule degradation. The proposed pre-equalization strategy again uses two types of molecules, and imitates destructive interference by considering the difference of type-A and type-B arrivals as the receiver molecule count. It is shown in the study that such an operation aggressively suppresses the ISI taps in the channel, greatly improving performance. It should, however, be noted that this performance enhancement requires the transmitted emitting type-B molecules with a certain delay and with a particular magnitude, which are both functions of channel parameters. Therefore, the strategy requires CSI (\textit{i.e.} the $f_{hit}(t)$ function, see Subsection \ref{subsec:estimation}) at the transmitter side. 

\par Furthermore, in \cite{typebasedsign_MNK}, the difference in the numbers of emitted type-A and type-B molecules is exploited to devise a modulation scheme, called the type-based sign (TS) modulation. Note that unlike molecular signals themselves (which are positive by nature), their differences can be negative. This property is exploited by \cite{typebasedsign_MNK} to mitigate the signal-dependent noise and to devise a transmitter side pre-equalization, similar to \cite{aminusb_burcu_2015}.That said, \cite{aminusb_burcu_2015} and \cite{typebasedsign_MNK} differ in a key aspect. In \cite{aminusb_burcu_2015}, type-A molecules carry the information and type-B (treated as the poison signal) follows type-A molecule emissions by mirroring them appropriately to reduce the ISI. On the other hand, in \cite{typebasedsign_MNK}, type-A and type-B molecules are used in a systematic way to provide positive and negative amplitudes (as is needed for the proposed scheme). In addition, the approach of \cite{typebasedsign_MNK} for cancelling the ISI term is through using a pre-coder for an effective channel, which is derived using reaction-diffusion equations. By considering the difference of two molecule types concentrations at the receiver, it is shown in \cite{typebasedsign_MNK} that the difference follows a linear differential equation that is not affected by the reaction rate between two molecule types. In \cite{typebasedsign_MNK}, it has also been shown that by making the two types react \textit{before} their arrival at the receiver, signal-dependent noise can be substantially reduced, leading to considerable improvements. We will demonstrate this effect in Subsection \ref{subsec:two_mols_BER}.

\subsection{Emission Time}
\label{subsec:emissiontime}

\par Using multiple types of molecules naturally incurs a higher transceiver complexity, which might be undesirable for nano- to micro-scale applications. As another alternative to emission intensity, we discuss emission time as a degree of freedom herein. 

\par DiMC systems that consider timing as the information source are considered in the form of two main strategies: continuous and discrete input alphabets. When considering continuous inputs, the propagation delay is treated as an additive noise \cite{first_MTC}. In a one-dimensional environment, the molecular timing channel (MTC) is generally referred to as the additive inverse Gaussian noise (AIGN) channel, and is studied in the context of its channel capacity \cite{AIGN_upperlower,MTC_capacitybound}. Note that in an MTC, instead of treating the emission times as continuous random variables, discretizing the timing input alphabet corresponds to the well-known pulse position modulation (PPM) (see the early review in \cite{cooke1947pulse}). PPM re-emerged in popularity in the 1990s with the introduction of ultrawideband radio concepts \cite{scholtz1997impulse,franz2006generalized,franz2007joint} for reasons that echo the interest in PPM for DiMCs. That is, the presence of significant multipath. The binary version of the PPM scheme ($2$-PPM) for a DiMC system is presented in Figure \ref{fig:modulation_diagram}. For a no-ISI channel, the optimal detector for PPM in a DiMC is given in \cite{murin_PPM_optimal}. The derived detector involves recording the arrival instants of every molecule, and performing relatively complex operations for detection. Motivated by this, lower complexity detectors are presented in \cite{murin_PPM_otherdetectors}.

\par For time-slotted digital DiMC systems \textit{with ISI}, pulse-position modulation (PPM) is initially considered in \cite{PPM_original_2011}. In the study, it is presented that BCSK yields better performance than the binary form of PPM (\textit{i.e.} $2$-PPM). We note that both schemes transmit at $t_{sym} = t_b$ as they are binary. However, each sub-slot of $2$-PPM is of length $\frac{t_b}{2}$, which results in increased ISI between consecutive sub-slots compared to BCSK. 

\par Recently, higher order PPM schemes have been considered in the context of DiMC \cite{bayram_PPM}. It is argued that for $K$-PPM, increasing $K$ results in a sparser transmission strategy over time, which is desirable for a DiMC system. Furthermore, as more bits are encoded in a single PPM symbol, each symbol emission can be made with a larger number of molecules while still satisfying the transmission power constraint presented in Section \ref{sec:LTI_poisson_section}. Combining these two beneficial attributes, it is shown in \cite{bayram_PPM} that higher order PPM schemes offer desirable performance improvements over conventional BCSK. 

\par We note that for the $K$-PPM, the time slots are assumed to equally spaced. Relaxing this consideration, \cite{timing_MNK} considers a $K$-ary time modulation scheme with $K$ unequally spaced releasing times and shows that this generalization outperforms the standard equal-interval timing modulation. In the study, the ML decision rule for this scheme is derived and is shown to have high complexity. Motivated by this, an efficient approach to simplify the ML decision rule to a simple threshold comparison is presented.

\par
\begin{figure}[!t] 
	\centering
	\includegraphics[width=0.48\textwidth]{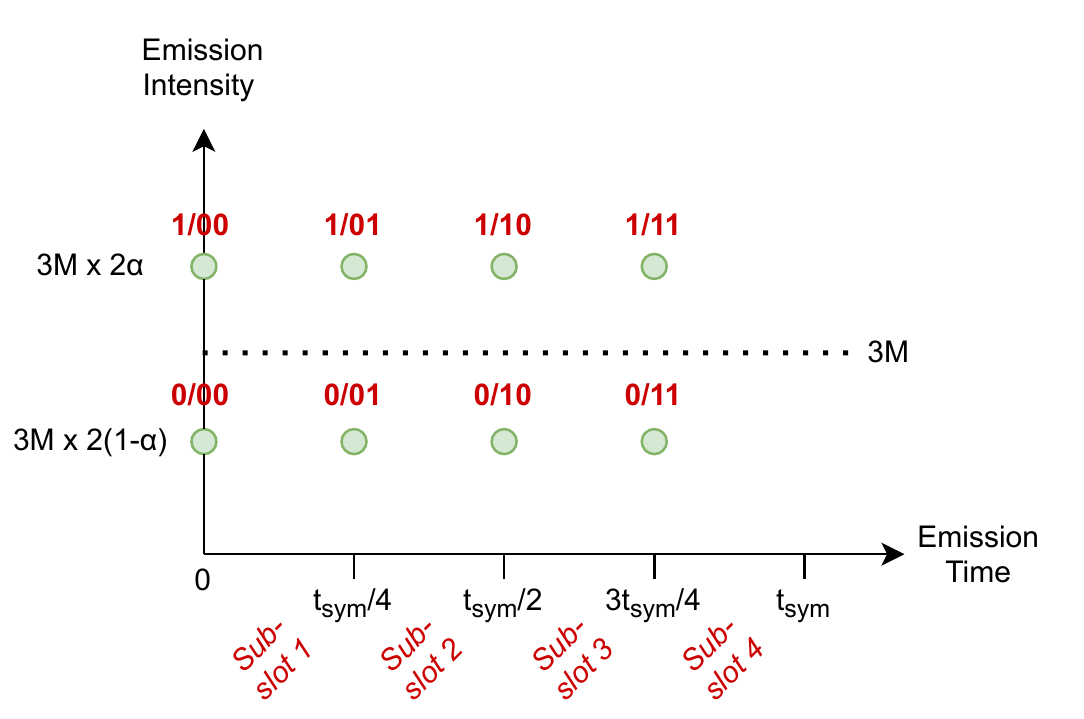} 
	\caption{A demonstrative two-dimensional constellation diagram of $4$-MCPM \cite{MCPM_mitra_2020}. Conventionally, the BCSK-bit can be considered to be the first bit in a ($1 + \log_2 K$)-bit symbol.}
	\label{fig:MCPM} 
\end{figure}

 In \cite{MCPM_mitra_2020}, emission intensity and time are jointly considered for information transfer. The proposed scheme is named the \textit{$K$-ary molecular concentration and position modulation} ($K$-MCPM), stemming from its building blocks: BCSK and $K$-PPM. As demonstrated in Figure \ref{fig:MCPM}, such a transmission strategy results in a two-dimensional constellation diagram. 
For a $K$-MCPM scheme, $B = 1 + \log_2 K$. Following the transmission power constraint, $K$-MCPM symbols can be transmitted by emitting $(1 + \log_2 K)M$ molecules on average. Thus, the concentration levels when the high and low BCSK-bits are transmitted with $2(1 + \log_2 K)M\alpha$ and $2(1 + \log_2 K)M (1-\alpha)$ molecules respectively, where $\alpha \in (0.5,1)$. Note that $\alpha \approx 0.5$ causes the concentration constellations to be hardly distinguishable. On the other hand, $\alpha \approx 1$ causes the PPM portion of MCPM to become hard to detect for the low BCSK-bit, since the emitted concentration is low. It is shown in \cite{MCPM_mitra_2020} that $\alpha$ governs a design trade-off for joint concentration and position modulations, and with proper optimization, MCPM can provide a promising error performance enhancement at higher data rates.

\subsection{Molecular Index Modulations}

\par Herein, we discuss signaling degrees of freedom other than the fundamental ones. Similar to its counterpart in traditional wireless communications \cite{index_general}, these methods can be grouped under molecular index modulations (molecular-IM). At its current stage of research, molecular-IM schemes can be grouped under two main categories: DiMC media-based modulations and DiMC spatial modulations.

\subsubsection{DiMC Media-Based Modulations}

\par In \cite{flowvelocity_MNK}, the molecular flow velocity meter approach is introduced. The key strategy in \cite{flowvelocity_MNK} is to design a transmitter-receiver pair that can measure the flow velocity in the DiMC channel. The study shows that this capability can be exploited to encode information using the flow velocity, leading to a new family of DiMC modulations. We note that as it transmits information using a physical property of the DiMC channel (rather than the emitted molecular signal), this new family resembles media-based modulations in RF communications \cite{mediabased_RF}. Furthermore, the study shows that the velocity meter approach is also capable of estimating the CSI in flow-assisted DiMC. 

\subsubsection{Molecular MIMO and DiMC Spatial Modulations}

\par
\begin{figure}[!t]
	\centering
	\includegraphics[width=0.4\textwidth]{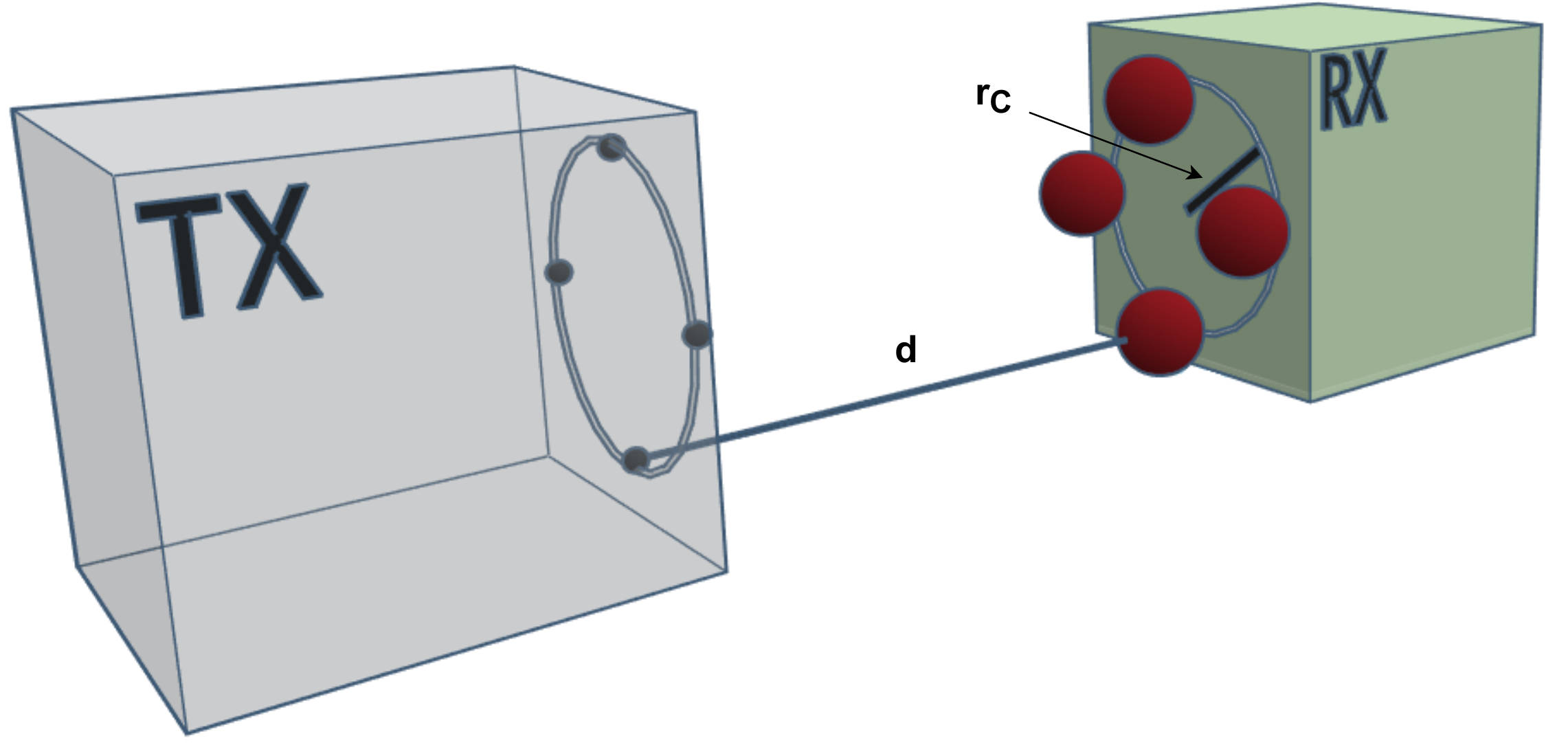} 
	\caption{A $4 \times 4$ molecular MIMO system. $d$ denotes the shortest distance between corresponding transmit and receiver antenna pairs and $r_C$ is the radius of the antenna array. This specific topology is referred to as the uniform circular array (UCA).}
	\label{fig:molecularMIMO}
\end{figure}

 Molecular MIMO aims to provide performance enhancement through introducing different emission sites on the transmitter body (\textit{i.e.} transmit antennas), and counting the molecule arrivals separately for different regions on the receiver body (\textit{i.e.} receiver antennas). Acknowledging the additional transceiver complexity it brings, molecular MIMO research is centered around how to best utilize these additional resources to improve communication performance. For reference, a $4 \times 4$ molecular MIMO system is presented in Figure \ref{fig:molecularMIMO}.

\par MIMO communications is first considered by \cite{MIMOfirst_akyildiz_2012} in the context of DiMC, through the discussion of basic spatial multiplexing (SMUX). The idea is more rigorously treated in \cite{MIMOtestbed_chae_2016}, where a physical macro-scale test-bed is also presented. It is shown therein that SMUX-based molecular MIMO faces high inter-link interference (ILI) in addition to ISI. The ILI is especially severe in SMUX that uses BCSK \cite{PPM_SMUX,estimationMIMO_spagnolini_2019}. To remedy this issue, \cite{estimationMIMO_spagnolini_2019} proposes to introduce \textit{time interleaving} and avoid simultaneous emissions from the transmitter antennas. In \cite{PPM_SMUX}, it is shown that a similar effect can also be achieved by using parallel streams of PPM instead of BCSK. 

\par In addition to SMUX-based studies, spatial diversity is considered in \cite{arraygain_damrath_2018}, where spatial repetition coding (same information is sent from all antennas) is shown to improve error performance over a SISO DiMC system. The improvement of this system is due to the increase in the total reception area of multiple receiver antennas. In addition, a single molecule-type, Alamouti-like space-time block code (STBC) \cite{alamouti} is also presented in the study. It is reported that since molecular signals are non-negative by nature, the orthogonality of the STBC is no longer maintained, and the results show that the scheme is outperformed by SISO BCSK. 

\par Recently, spatial modulation is considered as an alternative to SMUX-based MIMO approaches in RF communications, mainly due to the reduction it offers in transceiver complexity and its energy efficiency \cite{SM_original}. In essence, rather than employing diversity or SMUX, spatial modulation assigns indices to each of the transmit and receiver antenna pairs, and uses these indices to encode additional bits into the transmitted wave. In its original form, only one antenna is activated at a time. Therefore, an additional $\lfloor \log_2 (N_T) \rfloor$ bits can be encoded in the spatial constellations for an $N_T \times N_T$ MIMO setup. 

\par The spatial modulation concept is introduced to molecular MIMO systems in \cite{index_gursoy_2019,spatial_huang_2019}. The simplest form of DiMC spatial modulations, the molecular space shift keying (MSSK), encodes information solely into the antenna indices. At the receiver side, MSSK can be demodulated using a simple maximum count detector (MCD). For an $N_T \times N_T$ molecular MIMO setup with $N=1$, the MCD for MSSK can be formulated as
\begin{equation}
    \label{eq:MCD_MSSK}
    \hat{s}[k] = \mathop{\argmax}_{j \in \{1,\dots,N_T\}} \hspace{0.2cm} y^{(j)}[k],
\end{equation}
where $\boldsymbol{y}^{(j)}$ is arrival count vector at the $j^{th}$ receiver antenna. 

\par MSSK can also be combined with traditional schemes such as CSK \cite{spatial_huang_2019}, MoSK \cite{index_gursoy_2019}, and PPM \cite{PPMSM_gursoy_2019}, in order to jointly utilize both signal and spatial constellations for data transmission. It has been shown in these studies that DiMC spatial modulations outperform both diversity and SMUX-based schemes by a considerable margin, providing a promising performance enhancement over both SISO schemes and said MIMO approaches. Note that spatial diversity schemes do not combat ISI very efficiently, whereas SMUX provides aggressive ISI mitigation but suffers from severe ILI, caused by cross-talk between adjacent antenna pairs. Overall, a qualitative rule of thumb can be given as follows: DiMC spatial modulations provide better ISI mitigation than diversity schemes, but are generally worse than SMUX in this regard. However, the sparse antenna activation strategy provides a powerful ILI mitigation, which results in outperforming SMUX-based molecular MIMO. 

\par The MCD presented in \eqref{eq:MCD_MSSK} assumes that the first path is dominant. In other words, it assumes that a molecule emitted from transmit antenna $i$ is most likely to be received by the receiver antenna $i$. Thus, it works best when the antenna alignment is perfect, which is a strong assumption that cannot be guaranteed in nano- to micro-scale applications. For such cases, a simple equalization method is presented in \cite{misalignments_MIMO} to provide robustness against misalignments. That said, robustness against angular misalignments is still an open problem in molecular MIMO, alongside robustness against synchronization errors (see Section~\ref{subsec:synchronization}).

\section{DiMC Channel Capacity}
\label{sec:channel_capacity}

\par 
The information theoretic study of DiMC mainly centers around four channel models: the LTI-Poisson channel (intensity-based signaling) and the molecular timing channels (timing-based signaling) \cite{info_survey}, joint type-concentration channels \cite{zeroerror_MNK}, and joint concentration-time channels \cite{time_and_concentration}. %

\par Timing channels cover scenarios where the information transfer is through transmission/arrival times \cite{wray_timing_1992,moskowitz1992channel,STC_1994}. In such channels, the delay of an information can be considered analogously to an additive noise \cite{anantharam1996bits}. Thus, when considering timing channels in the context of DiMC (\textit{i.e.} MTCs), the additive noise is equivalent to the propagation delay a molecule exhibits \cite{first_MTC,AIGN}. Please note that as also mentioned in Subsection \ref{subsec:emissiontime}, since the information is encoded within the emission time, considered alphabet in an MTC is inherently continuous. For such channels, several works \cite{MTC_capacitybound,diffusion_based_MTC,inscribed_matter} study the fundamental limits of MTCs from an information theoretic sense, where the number of emitted molecules per transmission are considered one or many. Furthermore, we note that considering molecules with infinite lifetime also implies molecules that can have an arbitrarily large delay before arrival (see \cite{AIGN} for 1-D and Equation \eqref{eq:limitF} for 3-D). Motivated by this fact, these studies are extended by \cite{MTC_finitelifetime} where the particles have a finite lifetime. Overall, noting that this paper mainly considers intensity-based signaling, we shift our focus on the information theoretic studies regarding the LTI-Poisson model in the sequel. We point the reader to \cite{info_survey} for a more rigorous treatment of the MTC literature. 

\par The Poisson channel with an external noise source is considered in \cite{lapidoth_onthe,lapidoth_lowinput}, where upper and lower capacity bounds are provided. Furthermore, in ISI-free, emission intensity-based DiMC channels without external noise, (\textit{i.e., $L=1$, $N=1$, $\lambda_s = 0$}), the channel capacity is given in \cite{oneshot1}, alongside the capacity achieving input distribution. It is shown that given the transmitter can emit at most $M'$ molecules, the optimal distribution always has non-zero probability mass at $M'$ and $0$. A condition where the binary input distribution (\textit{i.e.}, OOK-based BCSK) becomes optimal is also provided. We also note that achievable rates for a finite-state Markov channel with feedback and CSI at the transmitter for the Poisson channel is provided in \cite{bacterialcable_capacity}, generalizing \cite{chen2005capacity}.  This work was motivated by behaviors of microbial communities which engage in electron transfer \cite{michelusi2016queuing,michelusi2014stochastic}.

\par In \cite{LTI_Poisson}, the more generalized LTI-Poisson model is considered, where $N=1$ and $L>1$. Note that this consideration involves ISI, hence the channel uses are not independent. Therefore, even though it is the case for the ISI-free scenario \cite{oneshot1}, the maximum mutual information for a single channel use does not fully characterize the LTI-Poisson capacity. For this peculiar case of $L>1$, several bounds on the capacity are derived in \cite{LTI_Poisson}. To the best of our knowledge, the exact capacity of the LTI-Poisson channel for $L>1$ is an open problem. 

\par In \cite{zeroerror_MNK}, molecule types are also considered as a degree of freedom, alongside emission intensity. The study considers the zero error capacity of a DiMC channel with finite memory length $L \geq 1$, when a finite number of molecule types is available at the transmitter. Depending on the maximum allowed emission intensity (of all types) or maximum number of molecules per each type, three coding schemes are proposed. In addition, lower and upper bounds on the zero error capacity are derived. The results show that as the number of available molecule types increases, the capacity of the system is substantially increases even though the total available molecules of all types are constrained.

\par To increase the information rate over emission-only or time-only signaling, joint concentration and time channels are considered in \cite{time_and_concentration}. While the release time can in general be continuous (similar to MTCs), the study considers a discrete value for the sake of practicality. Specifically, the transmission interval is divided into sub-intervals whereas the concentration is selected similar to previous LTI-Poisson studies. To determine the gain introduced by the additional degree of freedom, three achievable rates (capacity lower bounds) on the joint concentration-time channel are derived, considering three different detection schemes at the receiver side. Similar to the MTC literature, a no-ISI channel is assumed ($L=1$). Overall, the reported results suggest a considerable gain of jointly using concentration and time, over timing or intensity-based schemes alone.

\par The information theoretic study of DiMC channels provides insight into the fundamental limits of communication. That said, we will mainly focus on a DiMC system from design point of view. In the sequel, we discuss our perspective on the considerations in DiMC detector and equalizers from a design point of view. Starting from the optimal sequence detector,  we consider idealized scenarios which specialize to a conventional, fixed threshold detector and then discuss more sophisticated and higher performance state-of-the-art methods. Furthermore, we discuss the challenges of channel estimation and synchronization for DiMC systems. Specifically, Subsection \ref{subsec:estimation} considers DiMC channel estimation strategies alongside non-coherent detection schemes. The synchronization problem, clock mismatch estimation schemes, and asynchronous detection approaches are presented in Subsection \ref{subsec:synchronization}.

\section{Transceiver Signal Processing}
\label{sec:transceiver_SP}

\subsection{Detection and Equalization}
\label{subsec:detector}

\par Our metric of optimality is the  maximum  {\em a posteriori} probability.  Given the multipath induced by the DiMC, to achieve good performance, sequence detection will be necessary in many cases.  Under the assumption that each possible symbol is equally likely, this strategy reduces to maximum likelihood sequence detection (MLSD).  As famously shown in the seminal work,  complexity reduction is achievable if the memory/multipath is of finite duration and the Viterbi detector can be employed \cite{viterbi1967error}. In a DiMC system, the LTI-Poisson channel model clearly reveals the inherent multipath and channel memory. Recalling that the transmission vector $\boldsymbol{x}$ is related to $\boldsymbol{s}$ through the employed modulation scheme, and given data is transmitted in blocks with block length $S$, the MLSD operates by performing
\begin{equation}
    \label{eq:MLSD}
    \begin{split}
        \hat{\boldsymbol{s}} &= \argmax_{\boldsymbol{s}} P(\boldsymbol{y} | \boldsymbol{x}) \\
         &= \argmax_{\boldsymbol{s}} \prod_{n=1}^{S N} \frac{ \Tilde{\lambda}[n]^{y[n]}     e^{-\Tilde{\lambda}[n] } } {(y[n])!} \\
         &= \argmax_{\boldsymbol{s}} \sum_{n=1}^{S N} y[n] \ln (\Tilde{\lambda}[n]) - \Tilde{\lambda}[n].
    \end{split}
\end{equation}
Here, according to the candidate symbol sequence $\boldsymbol{s}$ and the channel model presented in Section \ref{sec:LTI_poisson_section}, the $SN$-vector $\boldsymbol{\Tilde{\lambda}}$ is expressed as
\begin{equation}
    \Tilde{\lambda}[n] = \sum_{n=1}^{S N} \bar{x}[n-k+1] h[n].
\end{equation}

\par We note that the MLSD presented in \eqref{eq:MLSD} is a generalized expression for an arbitrary (single-molecule) modulation scheme. That said, as mentioned in Section \ref{sec:LTI_poisson_section}, we focus on OOK-based BCSK as the default modulation scheme. For the special case where $L=1$ (no-ISI) and $N=1$ (single sample per bit), the maximum likelihood (ML) detector for BCSK is of the threshold form and can be written as 
\begin{equation}
    \label{eq:threshold_FTD}
    \hat{s}[k] = y[k] \mathop{\gtrless}_{0}^{1} \gamma.
\end{equation}
We refer to this specific scenario as the  {\em fixed threshold detector} (FTD) throughout the paper. For the no-ISI scenario, the optimal $\gamma$ value can be found by
\begin{equation}
    \gamma = \frac{2M h[1]}{\ln \big( \frac{2M h[1] + \lambda_s}{\lambda_s} \big)}.
\end{equation}

\par The FTD is a very simple detector with very low computational complexity, which is a desirable trait for a nano- or micro-scale transceiver. However, it is also fragile against ISI (even when $\gamma$ is optimized considering ISI \cite{ISI_burcu_2015}) and is sub-optimal for the case where $L>1$. Motivated by this, several studies consider adaptively changing the threshold and perform
\begin{equation}
    \label{eq:threshold_general}
    \hat{s}[k] = y[k] \mathop{\gtrless}_{0}^{1} \gamma[k]
\end{equation}
to decode the symbol. We call this family of strategies \textit{adaptive threshold detectors} (ATD) in this paper. The simplest form of ATDs selects $\gamma[k] = y[k-1]$ for $N=1$. For the general case where $N \geq 1$, it can be expressed as 
\begin{equation}
    \label{eq:ATD}
    \hat{s}[k] = \sum_{n = (k-1)N +1}^{kN} y[k] \mathop{\gtrless}_{0}^{1} \gamma[k]
\end{equation}
where $\gamma[k] = \sum_{n = (k-2)N +1}^{(k-1)N} y[k]$, which is equivalent to the energy collected within the previous symbol duration \cite{adaptivethreshold_damrath_2016}. This strategy is motivated by the fact that consequent symbols see comparable ISI energy, hence comparing the energy of consequent symbols provide information about the increase/decrease of concentration, hence the transmitted symbol. It is shown in \cite{adaptivethreshold_damrath_2016} that this detector outperforms FTD in higher data rates whilst falling short when the communication rate is slower. 

\par Note that the strategy of \cite{adaptivethreshold_damrath_2016} does not take channel properties into account. Using the channel coefficient vector $\boldsymbol{h}$, \cite{memorysamplingrate_mitra_2014} devises a more sophisticated ATD for BCSK,  denoted the \textit{memory-limited decision aided decoder} (MLDA). Given the strong ISI, the MLDA also employs decision feedback, that is we seek the following detector structure:
\begin{equation}
\label{eq:MLDFE}
\hat{s}[k] = \argmax_{j \in \{0, 1\}} P(\boldsymbol{y}_k | s[k] = j, \hat{s}[k-L],\cdots, \hat{s}[k-1]),
\end{equation}
where $\boldsymbol{y}_k = \begin{bmatrix} y[(k-1)N +1] & \dots & y[kN] \end{bmatrix}^T$. In MLDA, the current symbol is detected under the assumption that the previous $L$ symbols were correctly detected. We observe that a maximum \textit{a posteriori} sequence detector was also considered in \cite{MLSD_DFE_akan}. For $N=1$, \cite{memorysamplingrate_mitra_2014} shows that given perfect information about ISI and noise, the symbol-by-symbol maximum likelihood detector results in a threshold rule, where the threshold is given as:
\begin{equation}
    \label{eq:MLDA_1}
    \gamma[k] = \frac{2M h[1]}{\ln \big( \frac{2Mh[1] + \mathcal{I}[k] + N \lambda_s}{\mathcal{I}[k] + N \lambda_s} \big)}.
\end{equation}
Here, $\mathcal{I}[k]$ is the number of received molecules due to ISI. In practice, $\mathcal{I}[k]$ cannot be perfectly known, but can be estimated using past decisions and $\boldsymbol{h}$ by computing,
\begin{equation}
    \hat{\mathcal{I}}[k] = \sum_{i=2}^{L-1} h[i] \hat{x}[k-i+1].
\end{equation}
We note that due to complexity reasons, a nano-transceiver might not hold all $L$ previously detected symbols in memory. In that case, \cite{memorysamplingrate_mitra_2014} considers storing only $L' < L$ previously detected symbols. For the general case where $N \geq 1$ (\textit{i.e.} multiple samples per symbol), MLDA's symbol-by-symbol maximum likelihood detection becomes
\begin{equation}
    \label{MLDA_N}
     \sum_{q=1}^{N} y[(k-1)N+q] \ln \left(\frac{2M h[q] + \hat{\mathcal{I}}_q[k] + \lambda_s}{\hat{\mathcal{I}}_q[k]  + \lambda_s}\right) \mathop{\gtrless}_{0}^{1}   2M \sum_{q=1}^{N} h[q],
\end{equation}
where $\hat{\mathcal{I}}_q[k]$ is the estimated mean ISI contribution on the $q^{th}$ sample of the intended symbol.

\par MLDA provides strong ISI mitigation, hence an improved error performance. However, as it uses previous decisions for ISI estimation, it might be prone to error propagation when channel conditions are not favorable. Furthermore, it requires a more complex receiver than the conventional FTD, which might be undesirable for some applications. Motivated by this, rather than adaptively changing $\gamma[k]$ and keeping the transmission power constant for a bit-$1$ (\textit{i.e.} $2M$), \cite{ISI_burcu_2015,FTDadaptivetransmission_mitra_2016} consider the FTD at the receiving end, but vary the number of emitted molecules depending on earlier transmissions (\textit{i.e.} pre-equalization). By \cite{FTDadaptivetransmission_mitra_2016}, this strategy is called {\em adaptive transmission rate and constant thresholding} (ATRaCT). It should be noted that ATRaCT requires the availability of CSI (\textit{i.e.}, the vector $\boldsymbol{h}$) at the transmitter. That said, given CSI at the transmitter, ATRaCT is able to provide a strong performance improvement over FTD while still conserving the simplicity of the receiver nano-machine. Furthermore, it is shown in \cite{FTDadaptivetransmission_mitra_2016} that the scheme is more robust to distance variations than FTD, suggesting applicability when $d$ is not known exactly or is estimated erroneously.

\par Using the Viterbi decoder \cite{viterbi1967error}, the MLSD is of complexity $\mathcal{O}(2^L S)$. This exponential complexity can become prohibitive when the data rate is increased, as one needs to consider a larger $L$ to capture the same portion of $f_{hit}(t)$. Motivated by this concern, decision-feedback (DFE) and linear minimum mean squared error (linear MMSE, LMSSE) equalizers are derived in \cite{MLSD_DFE_akan}. We note that although they are classified under detectors, methods like MLDA also implicitly incorporate a DFE-type equalization within their set of operations. Using a similar approach, DFE is combined with Wald's sequential probability ratio test (SPRT, \cite{wald_SPRT}) in \cite{sprt_mitra} to devise a symbol-by-symbol detector. We will extend our discussion on SPRT approaches when we consider the impact of sychronization errors.
Overall, it can be inferred that despite its potential for error propagation, the DFE is a good and relatively low-complexity alternative to be paired with any type of DiMC detector.

\par Another new class of receiver side equalizers employ {\em discrete time derivative} pre-processing. First proposed by \cite{lin_derivative_2018}; therein it is observed that
applying a time derivative to the received sample vector $\boldsymbol{y}$ is shown to decrease the time dispersion of the signal. These initial endeavors are generalized for an arbitrary derivative order $m$ in \cite{gursoymitra_derivative_2020}, and it has been shown that higher order derivatives are able to outperform $m=1$ or no derivatives ($m=0$). However, it is also shown in \cite{gursoymitra_derivative_2020} that increasing $m$ arbitrarily does not lead to a monotonically improving error performance, as $m$ governs a trade-off between ISI mitigation and noise amplification. Herein, we will characterize said noise amplification phenomenon.

\begin{figure*}[!t]
	\centering
	\includegraphics[width=0.96\textwidth]{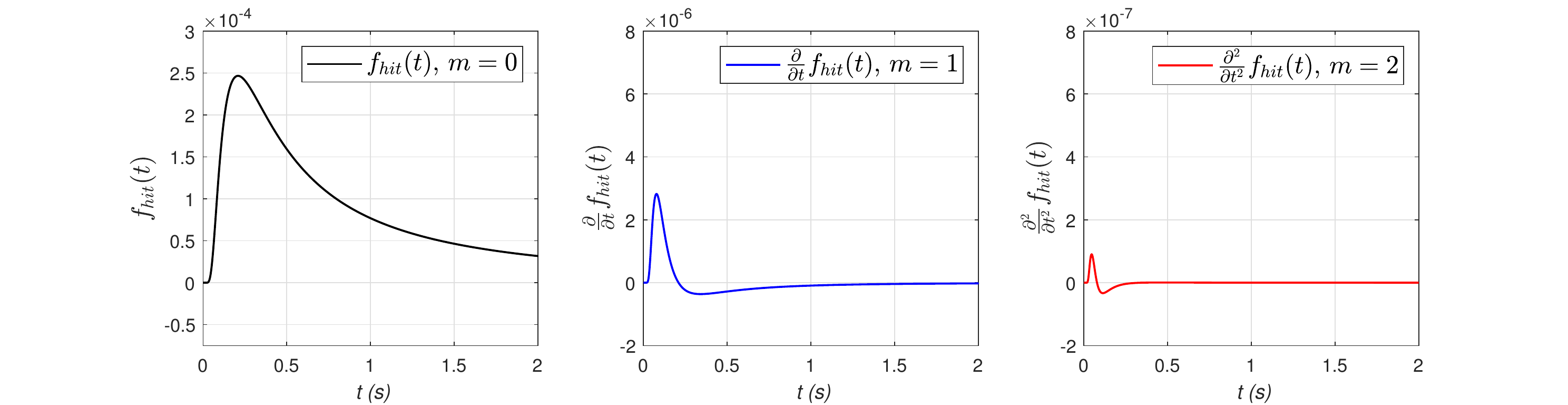} 
	\caption{$\frac{\partial^m f_{hit}(t)}{\partial t^m}$ for $m = 0$, $1$, and $2$ for $r_0 = \SI{15}{\micro\meter}$, $r_r = \SI{5}{\micro\meter}$, $D = 80 \frac{\SI{}{\micro\meter\squared}}{\SI{}{\second}}$. $m=0$ corresponds to the $f_{hit}(t)$ function itself in \eqref{eq:arrival_pdf}.}
	\label{fig:derivative_fhit}
\end{figure*}

\par Since the derivative operation is linear, it can be represented by a matrix $\boldsymbol{D}$, where 
\begin{equation} \label{eq:Dmatrix}
	\boldsymbol{D} = \begin{bmatrix}
		-1 & 1 & 0  & \cdots & 0 & 0\\
		0 & -1 & 1  & \cdots & 0 & 0\\
		\vdots & \vdots & \ddots & \vdots & \ddots & \vdots \\
		\vdots & \vdots &  & -1 & 1 & 0 \\
		0 & 0 & \cdots  & 0 & -1 & 1 \\
		0 & 0 & \cdots  & 0 & 0 & -1 \\
	\end{bmatrix}_{SN \times SN}.
\end{equation}
In addition, we can re-write the Gaussian approximation in \eqref{eq:y_seq_Gauss} in vector form as
\begin{equation} \label{eq:matrix_form}
	\begin{split}
    	\boldsymbol{y} = (\boldsymbol{H}\boldsymbol{x} + \lambda_s \boldsymbol{j}) + \boldsymbol{\eta},
	\end{split}
\end{equation}
where $\boldsymbol{j}$ is an vector of ones, $\boldsymbol{H}$ is a Toeplitz matrix of the coefficient vector $\boldsymbol{h}$, and $\boldsymbol{\eta} \sim \mathcal{N} (0,\textrm{diag}\{\boldsymbol{H}\boldsymbol{x}\} + \lambda_s \boldsymbol{I})$, following the definition in \eqref{eq:y_seq_Gauss}. Denoting $\textrm{diag}\{\boldsymbol{H}\boldsymbol{x}\} + \lambda_s \boldsymbol{I}$ as $\boldsymbol{\Sigma}$, applying the $m^{th}$ order derivative can be represented by 
\begin{equation} \label{eq:y_derivatived}
	\begin{split}
		\boldsymbol{y}_{(m)} &= \boldsymbol{D}^m \boldsymbol{y} \\
		&= \boldsymbol{D}^m \boldsymbol{H}\boldsymbol{x} + \boldsymbol{D}^m \lambda_s \boldsymbol{j} + \boldsymbol{D}^m \boldsymbol{\eta}. \\
		&= \boldsymbol{D}^m \boldsymbol{H}\boldsymbol{x} + \boldsymbol{D}^m \boldsymbol{\eta}.
	\end{split}
\end{equation}
Note that since $\eta \sim \boldsymbol{\Sigma}$, we have $\boldsymbol{D}^m \boldsymbol{\eta} \sim \mathcal{N} (0,\boldsymbol{D}^m \boldsymbol{\Sigma} (\boldsymbol{D}^T)^m ) $, which results in noise amplification. 

\par Overall, \cite{gursoymitra_derivative_2020} show that the $\boldsymbol{D}^m$ operator provides powerful ISI mitigation due to compression of the right-tail of the $f_{hit}(t)$ function. As a consequence of this, if the transceiver is able to transmit using large transmission powers, it can circumvent the noise amplification issue and achieve an order of magnitude higher data rates than conventional schemes while preserving reliable communication. In addition, the $\boldsymbol{D}^m$ operator provides:
\begin{enumerate}
    \item \textbf{Flexibility:} Note that the $m^{th}$ order derivative is an equalizer-like operator, and similar to the DFE, can be used as a separate block before the detection step. This makes operators like $\boldsymbol{D}^m$ and DFE compatible with different kinds of detectors, and provide flexibility in design.
    
    \item \textbf{A more feasible banded MLSD:} Recall that the MLSD is of complexity $\mathcal{O}(2^L S)$ using the Viterbi decoder. One might consider a sub-optimal banded MLSD, by considering a shorter memory $L' < L$, \cite{bandedMLSD}. Note that this version of the Viterbi decoder is of complexity $\mathcal{O}(2^{L'} S)$. However, in high data rate DiMC systems, this approach heavily underestimates ISI, resulting in poor error performance. That said, due to the aggressive ISI mitigation provided by the $\boldsymbol{D}^m$ operator, it is shown in \cite{gursoymitra_derivative_2020} that such a banded MLSD yields operable error performances. Therefore, the $\boldsymbol{D}^m$ operator makes MLSD-like detectors considerably more feasible for high data rate DiMC systems.
    
    \item \textbf{Very low computational complexity:} The $m^{th}$ order derivative operator is computationally very cheap. Even though its description is made through matrix multiplications for clarity of argument, essentially each derivative operation requires a single discrete time shift and a vector subtraction operation. Hence, recalling $S$ as the block length and $N$ as the samples per bit parameter, the $m^{th}$ order derivative operator is of complexity $\mathcal{O}(mSN)$ per block, which is linear in block length. Note that unlike DFE where the complexity is tied to the considered memory window, this complexity is regardless of the channel memory $L$.
\end{enumerate}

\subsection{On CSI Estimation and Transceiver Design}
\label{subsec:estimation}

\par Many detectors and equalizers mentioned in Section \ref{subsec:detector} rely on channel state information (CSI) at the receiver end, which is generally considered to be either the $f_{hit}(t)$ function itself, or the channel coefficient vector $\boldsymbol{h}$. In many studies throughout the literature, receiver CSI is assumed to be readily available, whereas in reality, the $\boldsymbol{h}$ vector needs to be estimated using practical algorithms \cite{survey_akan_2019}. Herein, we discuss recent endeavors in channel estimation for DiMC systems and non-coherent transceiver design strategies.

\par DiMC channel estimation is studied in \cite{estimation_noel_2015} in the context of joint parameter estimation, where the authors provide the Cramer-Rao Bound (CRB), ML estimation, and more practical estimators for jointly estimating $r_0$, $D$, molecule degradation rate, and number of emitted molecules for a single-shot emission. Note that estimating these parameters indirectly yields the $f_{hit}(t)$ function, if the form of $f_{hit}(t)$ is analytically known. However, the form is currently known for a small subset of topologies, under idealized conditions. Fortunately, for digital DiMC, the only information the transceivers need is the $\boldsymbol{h}$ vector to characterize the channel (assuming $\lambda_s = 0$ for the sake of discussion). Motivated by this, \cite{estimation_jamali_2016} proposes DiMC channel estimation strategies using pilot symbols. The CRB, ML estimation, and the linear least squares estimator are provided in the study, as well as strategies to select good pilot symbol sequences. In case the transceiver has access to prior information about $\boldsymbol{h}$, MAP estimation and corresponding Bayesian CRB are also derived. The endeavors of this study are generalized to molecular MIMO channels by \cite{estimationMIMO_spagnolini_2019}, where the estimation is done for channel coefficient matrix $\boldsymbol{H}_{\text{MIMO}}$. Note that the estimators of $\boldsymbol{H}_{\text{MIMO}}$ need to also take ILI.

\par As they rely on introducing pilot symbols in front of a data block, the strategies presented in \cite{estimation_jamali_2016} inherently reduce the effective data rate. Motivated by decreasing the number of pilot bits, the study is extended by \cite{semiblind_estimation_2020} where semi-blind channel estimation strategies are considered. In this context, expectation-maximization and decision-directed methods are proposed.

\par Instead of estimating $\boldsymbol{h}$ and employing the aforementioned detectors/equalizers in Subsection \ref{subsec:detector}, one may also consider non-coherent detectors that circumvent the need of channel estimation. For instance, some modulation schemes are compatible with simple $\argmax$ demodulators, \textit{i.e.} maximum count detectors (MCD). Examples include the basic type modulation (MoSK), pulse-position modulation (PPM), and the basic spatial modulation (MSSK \cite{index_gursoy_2019}). We note that since threshold optimization strategies require the CSI at the receiver end, the conventional demodulator of BCSK (FTD presented in \eqref{eq:threshold_FTD}) cannot be realized non-coherently. However, the simple ATD proposed in \cite{adaptivethreshold_damrath_2016} is capable of non-coherent operation, and might be compatible for simple nano- or micro-scale use due to its low complexity. In addition, neural network-based non-coherent detection is proposed in \cite{ANNdetection} for DiMC systems. It is shown in \cite{ANNdetection}, that the proposed detectors outperform Viterbi-based MLSD detection under imperfect CSI. That said, the method of\cite{ANNdetection} involves training a neural network, which can be a substantial effort, we note that their implementation is more suitable for macro-scale testbeds such as \cite{farsad_testbed,MIMOtestbed_chae_2016}.

\subsection{Synchronization}
\label{subsec:synchronization}

\par Like the availability of CSI, perfect synchronization between the transmitter and the receiver is also generally assumed for DiMC system design. However, achieving synchronization is a particularly challenging task in a DiMC system, especially considering the low molecule budget and limited computation capability of molecular transceivers. To this end, we discuss DiMC transceiver clock synchronization strategies and asynchronous receiver design herein.

\par In \cite{drift_lin_2017}, clock synchronization for a one-way DiMC with positive flow is considered. In the presented scheme, the messenger molecule is assumed to hold information about the transmission instant, which the receiver exploits to estimate the delay. In addition, a two-way message exchange is considered in \cite{AIGN_lin_2015,a_clock_linlin}. In \cite{AIGN_lin_2015}, delay and clock frequency imperfections are jointly considered. When the information molecule holds the transmission instant, it is shown by \cite{SIMO_sync} that clock recovery can be performed using a single symbol transmission in a single-input multiple-output (SIMO) system, using the arrival time differences at different antennas.

\par The above methods are inspired by conventional radio frequency-based (RF) communications, where the timing information is typically embedded within a packet. That said, encoding the transmission instant information within the molecule's chemical structure is a complex task for a nano-machine to handle. Therefore, it is of particular interest to discuss methods that do not assume such an encoding. In \cite{TNB_schober_2017,sync_twomolecules_2019}, the DiMC synchronization problem is tackled using two types of molecules: one for synchronization and the other for information transmission. For $K$-ary MoSK modulation, a blind synchronization approach (where the data is treated as a random variable) is presented in \cite{blind_2013}. A dual-molecule clock synchronization scheme for mobile molecular communication is also presented in \cite{mobile_sync}. 

\par Even without considering timing information within messenger molecule structure, the aforementioned synchronization methods rely on the use of multiple molecule types and or complex biochemical processes, which might be undesirable in terms of nano- to micro-scale transceiver complexity. Motivated by this, \cite{training_pcyeh} considers synchronization for single-molecule type intensity modulations. 

\par In addition to delay estimation approaches, the synchronization problem can also be tackled through developing asynchronous detection schemes. The simple asynchronous detector (ADS) and the asynchronous detector with decision feedback (ADDF) are presented in \cite{async_peak_noel}. For each BCSK symbol, ADS picks the sample with largest molecule count and compares the count with a fixed threshold, which can be written as
\begin{equation}
    \label{eq:ADS}
    \hat{s}[k] = \mathop{\max}_{q \in \{1, \dots, N \}} y[(k-1)N+q] \mathop{\gtrless}_{0}^{1} \gamma.
\end{equation}
Note that ADS resembles the conventional FTD presented in \eqref{eq:threshold_FTD}, with the main difference being in the considered received signal (\textit{i.e.} ADS considers the maximum sample, whereas \eqref{eq:threshold_FTD} considers the sum of samples). Using the same principle with an MLDA-like DFE mechanism, ADDF considers to mitigate ISI before employing ADS, which can be described by
\begin{equation}
    \label{eq:ADDF}
    \hat{s}[k] = \mathop{\max}_{q \in \{1, \dots, N \}} \big\{y[(k-1)N+q] - \hat{\mathcal{I}}_q[k] \big\} \mathop{\gtrless}_{0}^{1} \gamma,
\end{equation}
where the estimated ISI mean $\hat{\mathcal{I}}_q[k]$ can be found in a similar manner to MLDA.

\par We note that the maximum operations in \eqref{eq:ADS} and \eqref{eq:ADDF} do not \textit{predict} which sample holds the maximum according to the available CSI: they just take the maximum sample assuming the delay is $\tau = 0$. This way, even though the fixed-maximum sample method misses the peak for $\tau \neq 0$, ADS and ADDF are able to recover the maximum concentration as long as the peak stays within the symbol duration.  In \cite{async_peak_noel}, both ADS and ADDF are shown to outperform conventional schemes in terms of robustness against synchronization errors.

\par Another approach by which to increase robustness to synchronization errors is to adopt the principles of a {\em sequential probability ratio test} \cite{wald_SPRT}. In this scenario, the decision rule $\delta(\cdot)$ can be expressed as
\begin{align}
    \label{eq:DFE_SPRT}
    \delta(L_q(\mathbf{y}_q^k| \mathbf{x}))=\begin{cases}
    \hat{s}[k] = 0, &L_q(\mathbf{y}_q^k | \mathbf{x}) \leq A_{\text{SPRT}}\\
    \hat{s}[k] = 1, &L_q(\mathbf{y}_q^k | \mathbf{x}) \geq B_{\text{SPRT}}\\
    \mbox{sample}, &\mbox{else.}
    \end{cases}
\end{align}
where $q < N$, $L_q(\cdot)$ is the computed likelihood from the collected $\mathbf{y}_q^k$ vector, and $\mathbf{y}_q^k = \begin{bmatrix} y[(k-1)N+1] & \dots & y[(k-1)N+q] \end{bmatrix}^T$. 
The constants $A_{\text{SPRT}}$ and $B_{\text{SPRT}}$ are given by (as in \cite{wald_SPRT})
\begin{equation}
    A_{\text{SPRT}}=\frac{1-P_D}{1-P_{FA}}, \; B_{\text{SPRT}} = \frac{P_D}{P_{FA}}
    \label{eq:sprt_thresholds}
\end{equation}
where $P_{FA}$ is the false alarm rate and $P_D$ is the detection rate. The values of $P_{FA}$ and $P_{D}$ are set manually as targets. If a decision has not been made by $m=N$, a truncation rule based on the minimum distance rule is applied to prevent sampling into the subsequent channel symbol. In \cite{sprt_mitra}, the DFE-SPRT is proposed wherein the likelihood function is generalized to allow for decision feedback as in \eqref{eq:MLDFE}.

\par In \cite{sprt_mitra}, the DFE-SPRT is shown to provide strong robustness against synchronization errors, at the cost of a moderate increase in computational complexity over ADDF and MLDA. Furthermore, the synchronization error is estimated using a single molecule type by the maximum log-likelihood estimator. The CRB of the estimation is derived and the estimator is shown to yield close-to-CRB performance when the data rate is low. As described in \eqref{eq:DFE_SPRT}, the DFE-SPRT operates in iterations, where at each iteration, the receiver takes an additional sample and computes a likelihood ratio test (LRT) with the available vector of samples. Depending on the confidence, the symbol might be detected, or an additional sample is collected. We note that since the DiMC system is subject to ISI and the synchronization error estimation is non-perfect, the LRT is mismatched. It is shown in \cite{sprt_mitra} that under these imperfections, there exists an optimal number of samples and this number decreases with increasing mismatch. 

\par An interesting phenomenon is that for a DiMC system using BCSK and FTD, a small clock mismatch between the transmitter and the receiver actually improves error performance \cite{async_peak_noel,bayram_sync}. It is noted in \cite{bayram_sync} that the molecules that arrive within the first $\tau$ duration of the symbol interval are mostly due to ISI. Hence, assuming the receiver clock lags the transmitter clock by $\tau$, such a delay acts as an ISI mitigator and improves error performance. Of course, a large mis-synchronization results in bad performance, since the intended received signal power also decreases. It is shown in \cite{MCPM_mitra_2020} that this effect is applicable to PPM and MCPM as well. Furthermore, it is demonstrated in \cite{MCPM_mitra_2020} that higher order modulation schemes yield increased robustness against more severe synchronization errors (\textit{i.e.} large $\tau$). We note that while satisfying the same bit duration ($t_b$) constraint, a higher order modulation scheme is able to transmit at a larger symbol duration (see Section \ref{sec:LTI_poisson_section}). This makes $\tau$ comparatively smaller with respect to $t_{sym}$, alleviating the debilitating effect of the mismatch.

\par Overall, the synchronization problem is a particularly challenging one for nano- to micro-scale DiMC systems, due to their low-complexity requirements and the DiMC channels' properties. Although the sub-field is still relatively less explored, we believe that detector and equalizer designs that seek to provide robustness against synchronization errors may have greater practical utility than those that do not.  This remains an open avenue for research.
In addition, single-molecule type clock recovery strategies can be devised, as they are in line with the low-complexity requirements of DiMC transceivers.

\section{Coding for DiMC}
\label{sec:channel_coding}

\par Error control and source coding in DiMC systems are particularly interesting problems due to the signal-dependent arrival statistics and ISI. The peculiarities of the DiMC channel in terms of error control code design can be grouped under two main groups:
\begin{enumerate}
    \item \textbf{Asymmetric codeword distances:} Since received signal statistics change according to previous emissions, the distances between codeword pairs are asymmetric. Therefore, the regular Hamming distance does not fully capture the error probabilities between codewords. 
    \item \textbf{Code rate and ISI:} For an ($n,k$) code, a code rate $r = \frac{k}{n}$ that is lower than one increases ISI. For example, consider $r = \frac{3}{5}$. In order to satisfy an information bit duration constraint of $t_b$, each codeword bit can only be allocated $\frac{3t_b}{5}$ duration per transmission. This poses a trade-off between error control capability and ISI amplification avoidance, and suggests that higher rate codes are more favorable for DiMC systems. Also note that the transmission power normalization suggests a similar reduction in transmission power as well.
\end{enumerate}

\par Classical Hamming codes are considered in \cite{hamming_leeson_2012} in the context of DiMC, whereas classical convolutional codes are utilized in \cite{convolutional1}. A code family under the name {\em ISI-free codes} is proposed in \cite{ISIfree_2013}, and self-orthogonal convolutional codes are presented to the DiMC domain by \cite{SOCC_2015}. Reed-Solomon codes are proposed to DiMC systems with multiple transmitters and a single receiver in \cite{RScodes_mitra_2019}. 

\par In a DiMC system that uses OOK-BCSK, the sources of ISI are the residual molecules coming from bit-$1$ transmissions. Thus, a sparser transmission strategy generates less ISI and improves error performance. Using this principle, \cite{kislal_2020} considers a heuristic code construction. In addition, due to the transmission power constraint, using a sparser code translates to having a larger transmission power for each bit-$1$ transmission in the codeword. Motivated by this, \cite{MEC_2014} propose the minimum energy code (MEC) family. 


\par Recall from the first bullet point that due to ISI, the distances between codeword pairs are asymmetric in DiMC. Motivated by this fact, \cite{moco_distance_2012} proposes to use a distance measure specific to DiMC channels (the molecular coding distance function,  {\em MoCo distance}). The  {\em MoCo distance} is defined as
\begin{equation}
    \label{eq:MoCo_distance}
    d(\boldsymbol{c}_i,\boldsymbol{c}_j) = - \log P(\boldsymbol{c}_i \rightarrow \boldsymbol{c}_j),
\end{equation}
where $P(\boldsymbol{c}_i \rightarrow \boldsymbol{c}_j)$ denotes the probability of decoding codeword $j$ when codeword $i$ is sent. A similar approach is also taken in \cite{ISIaware_oguz_2019}, where codebook construction is based on codeword distances in a transition probability matrix. The results of the study show that the ISI enhancement due to having $r < 1$ causes many traditional coding strategies to be outperformed by uncoded BCSK, emphasizing that the error correction capability of a DiMC code design needs to overcome the increased ISI. 

\par Error control coding literature for DiMC is still in its infancy, with many open problems ahead. Herein, we list several design considerations that we believe will aid future research in this subfield:
\begin{enumerate}
    \item {\bf New distance metrics:} Traditional Hamming distance is not an ideal metric in DiMC channels due to the aforementioned channel characteristics. However, directly using transition probabilities between codewords heavily depend on channel statistics, which is a function of $r_r$, $r_0$, $D$, and $t_b$. Thus, a possible shortcoming of transition probability-based code designs is the lack of generalizability, which can become a major issue if the transmitter and/or receiver are mobile, $D$ changes over time, or synchronization is imperfect. Code designs for DiMC need to be robust against these unwanted changes. In fact, robustness against these fluctuations can itself be a design objective for code construction.
    
    \item {\bf Low complexity methods for high data rates:} Using transition probability matrix entries as objective functions for an ($n,k$) block code construction implies that the matrix of interest is of size $2^n \times 2^n$, even when assuming no spill-over ISI from previous codeword transmissions \cite{ISIaware_oguz_2019}. However, pushing for higher data rates incurs a longer ISI window, necessitating a codeword memory window consideration of $L_c > 1$ for accurate designs. Unfortunately, this causes the transition matrix to become $(2^n)^{L_c} \times (2^n)^{L_c}$, which becomes computationally expensive for larger codeword lengths. In this regard, similar to the signal-to-interference and noise ratio (SINR)-BER relationship in traditional wireless communications, devising a proxy objective function for code design can be highly beneficial for future research.
    
    \item {\bf Coded-modulation designs:} Recalling its description in Subsection \ref{subsec:type_modulations}, given a channel memory $L$, the ISI avoiding modulation approach avoids emitting the same type of molecules within $L$ consecutive symbol durations. Naturally, this approach leads to constraining the channel input sequences that are compatible with the modulation. Therefore, even though they are described as modulation schemes, ISI avoiding modulations inherently incorporate source coding within their strategy. That said, to the best of our knowledge, other than several studies including \cite{zeroerror_MNK,ISIavoiding_MNK}, the error control coding literature for DiMC mostly assumes solely emission intensity-based modulations as the employed modulation scheme. However, it should be noted that since the modulation design heavily affects the observed ISI, \textit{the efficacy of DiMC coding strategies is coupled with the employed modulation scheme.} Therefore, joint code-modulation designs can leverage the degrees of freedom in both fields, and develop a more powerful and adaptive strategy. 
    
\end{enumerate}

\section{Numerical Results}
\label{sec:PerformanceEval}

\par Herein, we compare the detectors and several modulation schemes discussed in the paper, with a goal of providing intuition regarding their strengths and weaknesses in varying scenarios. Throughout the section, unless stated otherwise, uncoded OOK-based BCSK is employed as the default modulation scheme. In the section, signal-to-noise ratio (SNR) defines the ratio of the mean transmission power ($M$) and the external Poisson noise rate per bit ($N \lambda_s$), as
\begin{equation}
    \label{eq:SNR}
    \textrm{SNR} = \frac{M}{N \lambda_s}.
\end{equation}
Furthermore, the $m^{th}$ order derivative operators are evaluated using the ADS detector to provide a very low complexity receiver design. It is worth noting that the last $m$ samples of each symbol is discarded for the $\boldsymbol{D}^m$-ADS scheme, in order to avoid the non-causal ISI introduced by the forward derivative operator \cite{lin_derivative_2018,gursoymitra_derivative_2020}. Specifically, the following operation is performed for detection:
\begin{equation}
    \label{eq:ADS_Dm}
    \hat{s}[k] = \mathop{\max}_{q \in \{1, \dots, N-m \}} y_{(m)}[(k-1)N+q] \mathop{\gtrless}_{0}^{1} \gamma.
\end{equation}

\subsection{Error Performances of Detectors}
\label{subsec:BER_detectors}

\par
\begin{figure}[!t]
	\centering
	\includegraphics[width=0.48\textwidth]{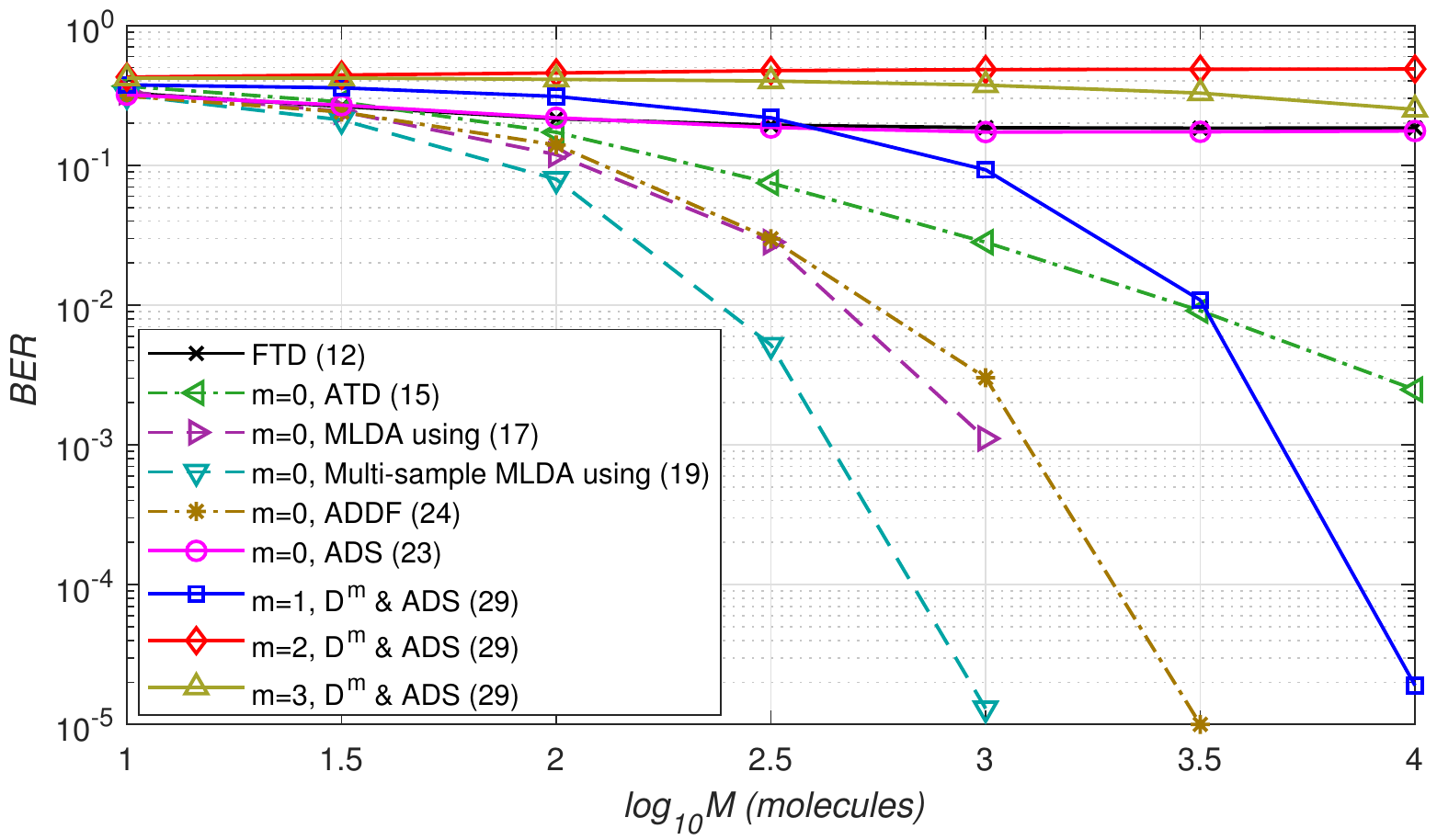}
	\caption{BER vs. $M$ curves for various detection strategies. $r_0 = \SI{10}{\micro\meter}$, $r_r = \SI{5}{\micro\meter}$, $D = 80 \frac{\SI{}{\micro\meter\squared}}{\SI{}{\second}}$, data rate $R = 10$ bits per second ($t_b = \SI{0.1}{\second}$), $\textrm{SNR} = 10$dB, $L=40$, $S=1000$, $N=5$, $\tau = 0$. Considered decision feedback memory $L' = 30$ for ADDF and MLDA, $\gamma$ values numerically optimized for all methods.}
	\label{fig:20bps_comparison}
\end{figure}

\begin{figure}[!t]
	\centering
	\includegraphics[width=0.48\textwidth]{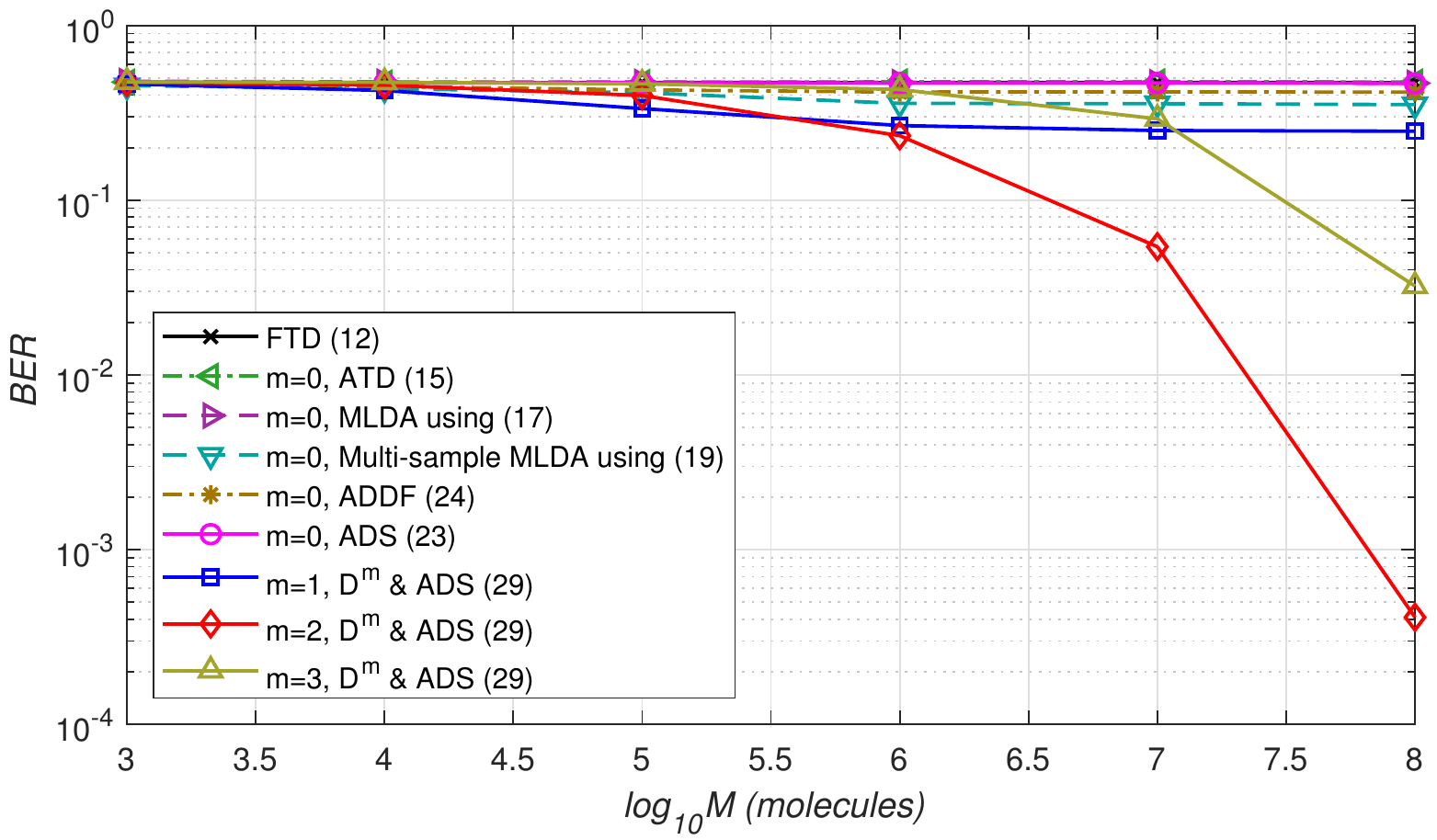} 
	\caption{BER vs. $M$ curves for various detection strategies. $r_0 = \SI{10}{\micro\meter}$, $r_r = \SI{5}{\micro\meter}$, $D = 80 \frac{\SI{}{\micro\meter\squared}}{\SI{}{\second}}$, data rate $R = 50$ bits per second ($t_b = \SI{0.02}{\second}$), $\textrm{SNR} = 10$dB, $L=200$, $S=1000$, $N=5$, $\tau = 0$. Considered decision feedback memory $L' = 150$ for ADDF and MLDA, $\gamma$ values numerically optimized for all methods.}
	\label{fig:50bps_comparison}
\end{figure}

 Firstly, we present error performances of selected detectors presented in Section \ref{sec:transceiver_SP}. Using the aforementioned considerations, Figures \ref{fig:20bps_comparison} and \ref{fig:50bps_comparison} are presented to demonstrate the comparative BER performances of the detection schemes under lower and higher data rates, respectively.

\par The results of Figures \ref{fig:20bps_comparison} and \ref{fig:50bps_comparison} show that different detector strategies have the edge at different data rates. For relatively lower data rates \textit{i.e.} Figure \ref{fig:20bps_comparison}, it can be observed that MLDA and ADDF provide strong advantages over the rest of the evaluated schemes. Note that FTD in \eqref{eq:threshold_FTD}, ATD in \eqref{eq:ATD}, ADS in \eqref{eq:ADS}, and $\boldsymbol{D}^m$ \& ADS in \eqref{eq:ADS_Dm} are very low complexity approaches that do not estimate ISI. Naturally, as MLDA and ADDF incorporate ISI estimation/mitigation strategies within their set of operations, they outperform the less complex methods.

\par Interestingly, Figure \ref{fig:50bps_comparison} shows that the superiority of MLDA and ADDF does not continue as the data rate increases. Note that an increase in data rate corresponds to a considerably higher ISI, causing these methods to be more prone to error propagation as $t_b$ decreases. At the high data rate regime, higher order derivative schemes ($\boldsymbol{D}^m$ \& ADS where $m=2$ and $3$) are observed to provide the best error performance among the evaluated schemes. This advantage is due to the aggressive ISI mitigation offered by the $\boldsymbol{D}^m$ operators. Note that ISI mitigation is stronger for $m>1$ compared to $m=1$, which can also be confirmed from Figure \ref{fig:derivative_fhit}. That said, Figure \ref{fig:50bps_comparison} shows that the error performance does not monotonically improve with $m$. This phenomenon is due to the noise amplification issue described in Subsection \ref{subsec:detector}. Recalling that noise amplification becomes more severe with increasing $m$, the trade-off between ISI mitigation and noise amplification avoidance suggests that there exists an optimal derivative order that minimizes BER (which is a function of the system parameters) \cite{gursoymitra_derivative_2020}. Overall, using the proper $m$ value (\textit{e.g.} $m=2$ for Figure \ref{fig:50bps_comparison}) and given the transmitter can handle large transmission powers to overcome noise amplification\footnote{We note that as the arrivals are Poisson random variables, their mean over standard deviation ratios improve with increasing intensity.}, a reliable error performance can be preserved while significantly increasing the data rate. Furthermore, this increase can be achieved with a very simple detector, which we believe is promising for nano- to micro-scale applications.

\subsection{Robustness Against Synchronization Errors}
\label{subsec:BER_synchronization}

\par
\begin{figure}[!t]
	\centering
	\includegraphics[width=0.48\textwidth]{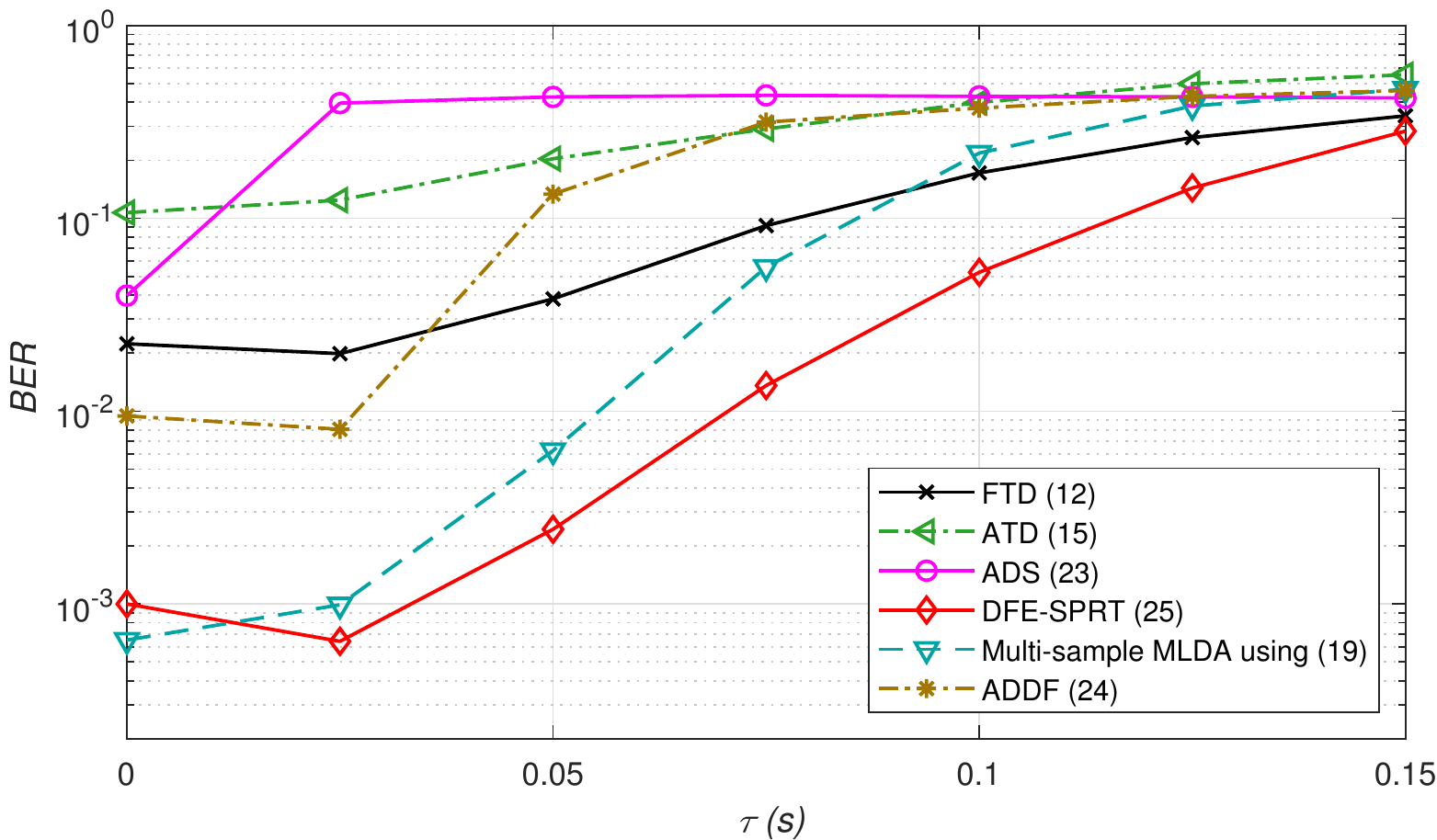} 
	\caption{BER vs. $\tau$ for various detection strategies. $M = 10^{2}$ molecules, $r_0 = \SI{10}{\micro\meter}$, $r_r = \SI{5}{\micro\meter}$, $D = 80 \frac{\SI{}{\micro\meter\squared}}{\SI{}{\second}}$, data rate $R = 4$ bits per second ($t_b = \SI{0.25}{\second}$), $\textrm{SNR} = 20$dB, $L=16$, $S=1000$, $N=5$. Considered decision feedback memory $L' = L$ for all applicable schemes. $\gamma$ values numerically optimized for $\hat{\tau} = 0$.}
	\label{fig:4bps_SPRT}
\end{figure}

\begin{figure}[!t]
	\centering
	\includegraphics[width=0.48\textwidth]{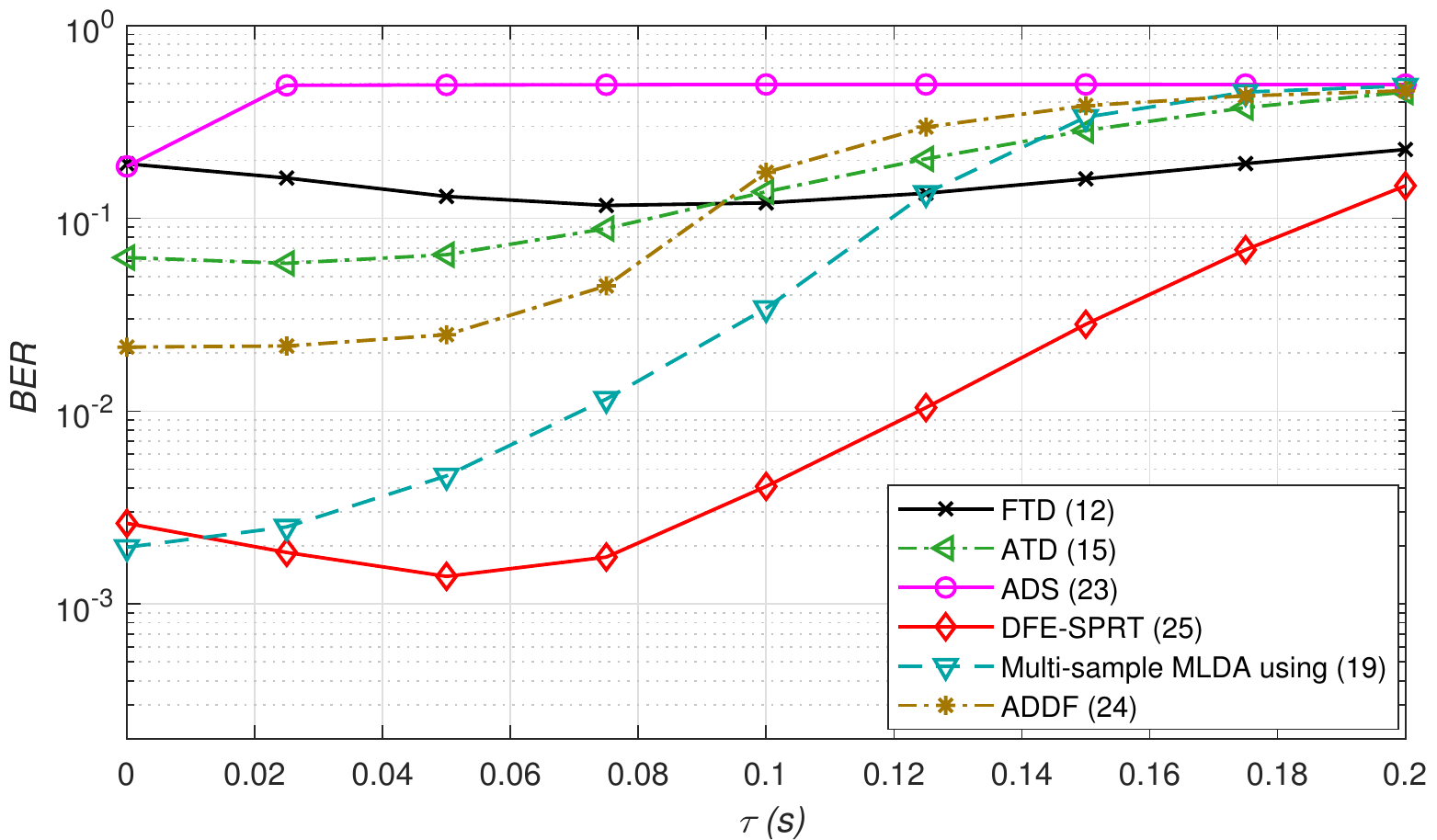} 
	\caption{BER vs. $\tau$ for various detection strategies. $M = 10^{2.5}$ molecules, $r_0 = \SI{10}{\micro\meter}$, $r_r = \SI{5}{\micro\meter}$, $D = 80 \frac{\SI{}{\micro\meter\squared}}{\SI{}{\second}}$, data rate $R = 10$ bits per second ($t_b = \SI{0.1}{\second}$), $\textrm{SNR} = 20$dB, $L=40$, $S=1000$, $N=5$. Considered decision feedback memory $L' = L$ for all applicable schemes. $\gamma$ values numerically optimized for $\hat{\tau} = 0$.}
	\label{fig:10bps_SPRT}
\end{figure}

 In addition to the error performance under perfect synchronization, robustness to synchronization errors is also of particular interest. In order to demonstrate the robustness of select schemes, Figures \ref{fig:4bps_SPRT} and \ref{fig:10bps_SPRT} are presented to evaluate BER vs. $\tau$ for low and relatively higher data rates, respectively. The schemes in Figure \ref{fig:4bps_SPRT} and \ref{fig:10bps_SPRT} employ no clock recovery strategy, in order solely evaluate the robustness of the considered detectors. In other words, all methods assume that the receiver is perfectly synchronized, $\hat{\tau} = 0$, and have access to CSI for the synchronized case. 

\par The results of Figures \ref{fig:4bps_SPRT} and \ref{fig:10bps_SPRT} show that DFE-SPRT provides the best robustness among evaluated schemes in both milder and harsher channels. Furthermore, even though the comparative performance rankings of the evaluated methods mostly stay the same, the results of Figure \ref{fig:10bps_SPRT} show an interesting phenomenon: A small amount of delay might improve the error performance for schemes in higher ISI scenarios. As also mentioned in Subsection \ref{subsec:synchronization}, this phenomenon is also reported in \cite{async_peak_noel,bayram_sync}. To explain this phenomenon, we recall that the intended symbol's molecules experience a propagation delay before arriving at the receiver. Therefore, the molecules that arrive at the beginning of the symbol duration are more likely to be residual molecules from previous transmissions, rather than the ones carrying the intended symbol. The receiver that lags the transmitter clock by $\tau$ is able to discard these residual molecules, effectively mitigating ISI. 

\subsection{MLDA vs. ATRaCT: Receiver vs. Transmitter Side Equalization}

\par
\begin{figure}[!t]
	\centering
	\includegraphics[width=0.48\textwidth]{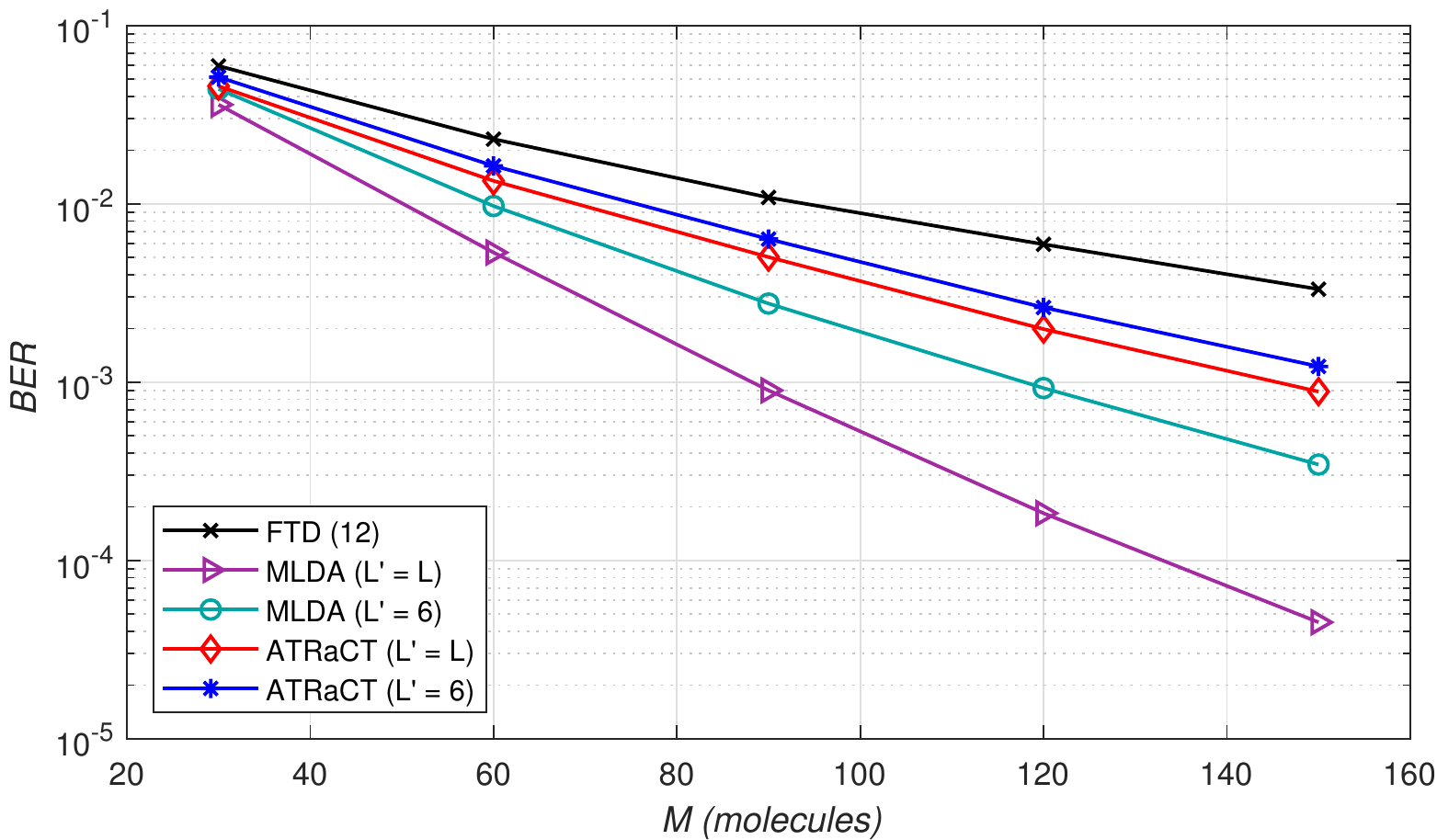} 
	\caption{BER vs. $M$ for FTD, MLDA, and ATRaCT \cite{FTDadaptivetransmission_mitra_2016}. $r_0 = \SI{10}{\micro\meter}$, $r_r = \SI{5}{\micro\meter}$, $D = 80 \frac{\SI{}{\micro\meter\squared}}{\SI{}{\second}}$, data rate $R = 3$ bits per second ($t_b = \frac{1}{3}\SI{}{\second}$), $\textrm{SNR} = 10$dB, $L=12$, $S=1000$, $N=1$, $\tau = 0$. $\gamma$ values numerically optimized for FTD and ATRaCT. Average transmission power of ATRaCT is normalized to $M$ for fairness.}
	\label{fig:3bps_ATRACT}
\end{figure}

 As shown in the previous subsections, MLDA is a powerful detection scheme, stemming from its effective use of CSI to adaptively change the threshold value (see \eqref{eq:MLDA_1}). That said, MLDA also brings more computational burden to the receiver. As also discussed in Section \ref{subsec:detector}, ATRaCT moves the equalization block of MLDA to the transmitter as a \em{pre-equalizer} or \em{pre-coder}, and considers keeping the simple FTD at the receiver while still controlling the ISI. Herein, we demonstrate their comparative performances in Figure \ref{fig:3bps_ATRACT}, and their gains over the conventional FTD. 

\par The results of Figure \ref{fig:3bps_ATRACT} show that ATRaCT is outperformed by MLDA. ATRaCT does indeed outperform FTD-based BCSK, which is due to its inherent ISI mitigation. Overall, it is argued in \cite{FTDadaptivetransmission_mitra_2016} that the use of ATRaCT is dependent on the application requirements. It is exemplified therein that as MLDA in \eqref{eq:MLDA_1} calculates logarithms, it requires more computational resources than a simple threshold comparison, at the cost of a loss in error performance. As a rule of thumb, we note the advantage of ATRaCT when the receiver structure is bound to be simplistic, whereas for more capable and complex receivers, MLDA has the edge. That said, it should be noted that an adaptive receiver scheme can also be used in conjunction to the adaptive transmitter, further improving performance \cite{FTDadaptivetransmission_mitra_2016}.

\subsection{Low-Complexity and Low-Transmission Power Modulations}
\label{subsec:BER_MCPM}

\par Depending on the applications, it might be of particular interest to establish nano- or micro-scale DiMC links that require very low transceiver complexity. For these applications, using multiple molecule types or considering sophisticated detection and equalization strategies become undesirable. Using the results of Figure \ref{fig:50bps_comparison}, we had noted in Subsection \ref{subsec:BER_detectors} that if the simple transceiver has access to a large storage of molecules, derivative-based pre-processing is able to offer a desirable ISI mitigation for BCSK, using a very simple set of operations. However, due to its noise amplification, this method yields high error probabilities for low transmission powers. 

\par
\begin{figure}[!t]
	\centering
	\includegraphics[width=0.48\textwidth]{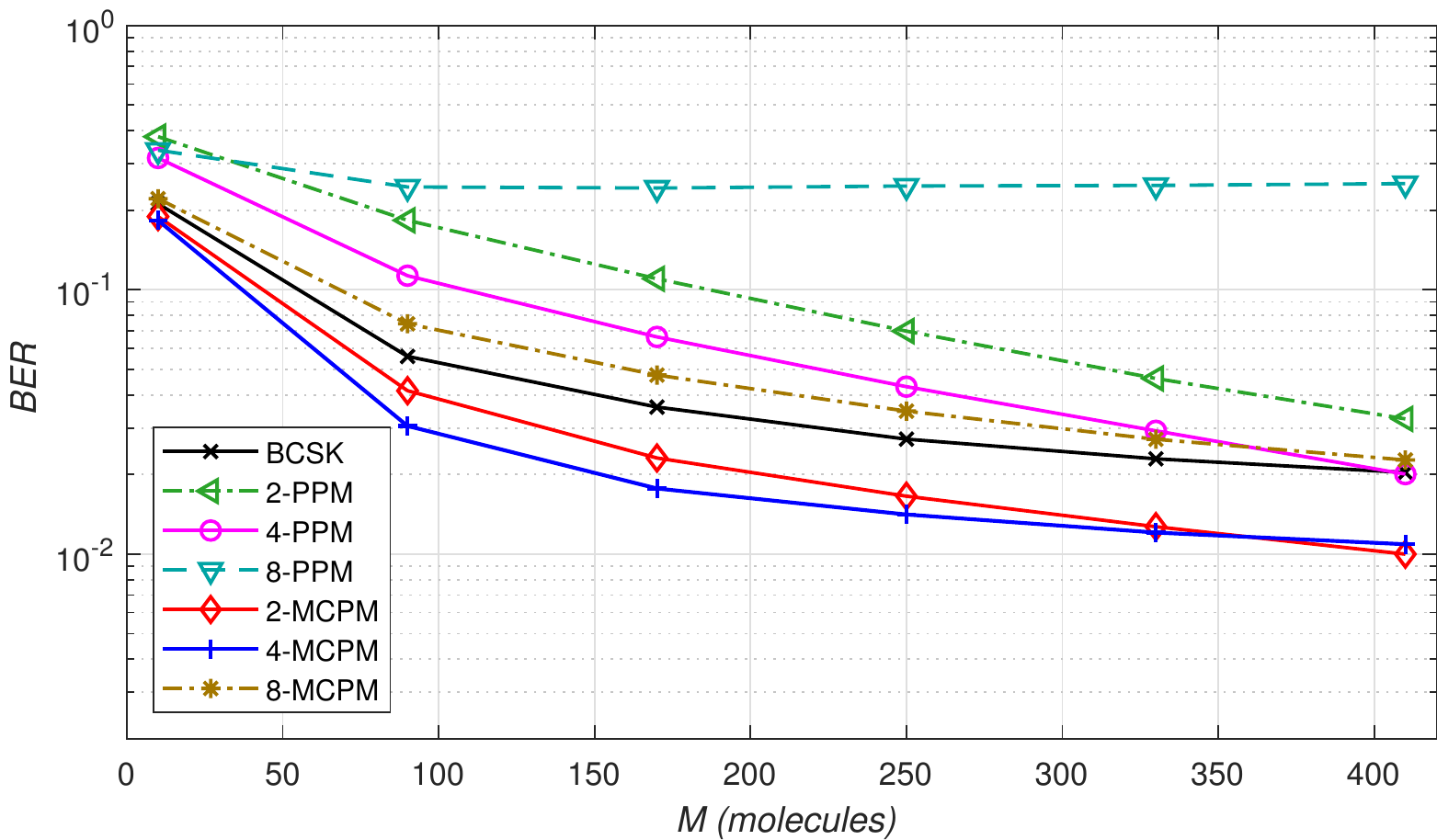} 
	\caption{BER vs. $M$ for various detection strategies \cite{MCPM_mitra_2020}. $r_0 = \SI{10}{\micro\meter}$, $r_r = \SI{5}{\micro\meter}$, $D = 80 \frac{\SI{}{\micro\meter\squared}}{\SI{}{\second}}$, data rate $R = 5$ bits per second ($t_b = \SI{0.2}{\second}$), $\textrm{SNR} = 20$dB, $L=20$, $S=9\times 10^4$, $N=1$, $\tau = 0$. $\alpha$ and $\gamma$ values numerically optimized for MCPM and BCSK.}
	\label{fig:5bps_MCPM}
\end{figure}

 In Figure \ref{fig:5bps_MCPM}, we present error performance results for the modulation design problem for low-complexity and low-transmission power DiMC modulations. We specifically consider single-molecule type schemes, and SISO transmission. Referring to the discussion in Section \ref{sec:degrees_of_freedom}, under these considerations, we specifically consider emission intensity (BCSK), emission time (PPM), and joint intensity-time (MCPM) modulations for comparison. Acknowledging the complexity considerations, BCSK is demodulated using the FTD in \eqref{eq:threshold_FTD}, PPMs are demodulated using the $\argmax$-based maximum count detector (MCD), and MCPMs are demodulated using a two stage detector that first employs MCD for PPM detection, and FTD for concentration detection. 

\par The results of Figure \ref{fig:5bps_MCPM} show that the joint use of concentration and time outperform intensity-only or time-only schemes in the regime of interest. Recalling the transmission power and bit duration normalizations, the key to MCPM schemes' success is that they are able to encode more bits within a symbol while still exploiting the temporal sparsity of a PPM-like transmission strategy.

\subsection{Second Molecule Type: How to Best Utilize it?}
\label{subsec:two_mols_BER}

\par
\begin{figure}[!t]
	\centering
	\includegraphics[width=0.48\textwidth]{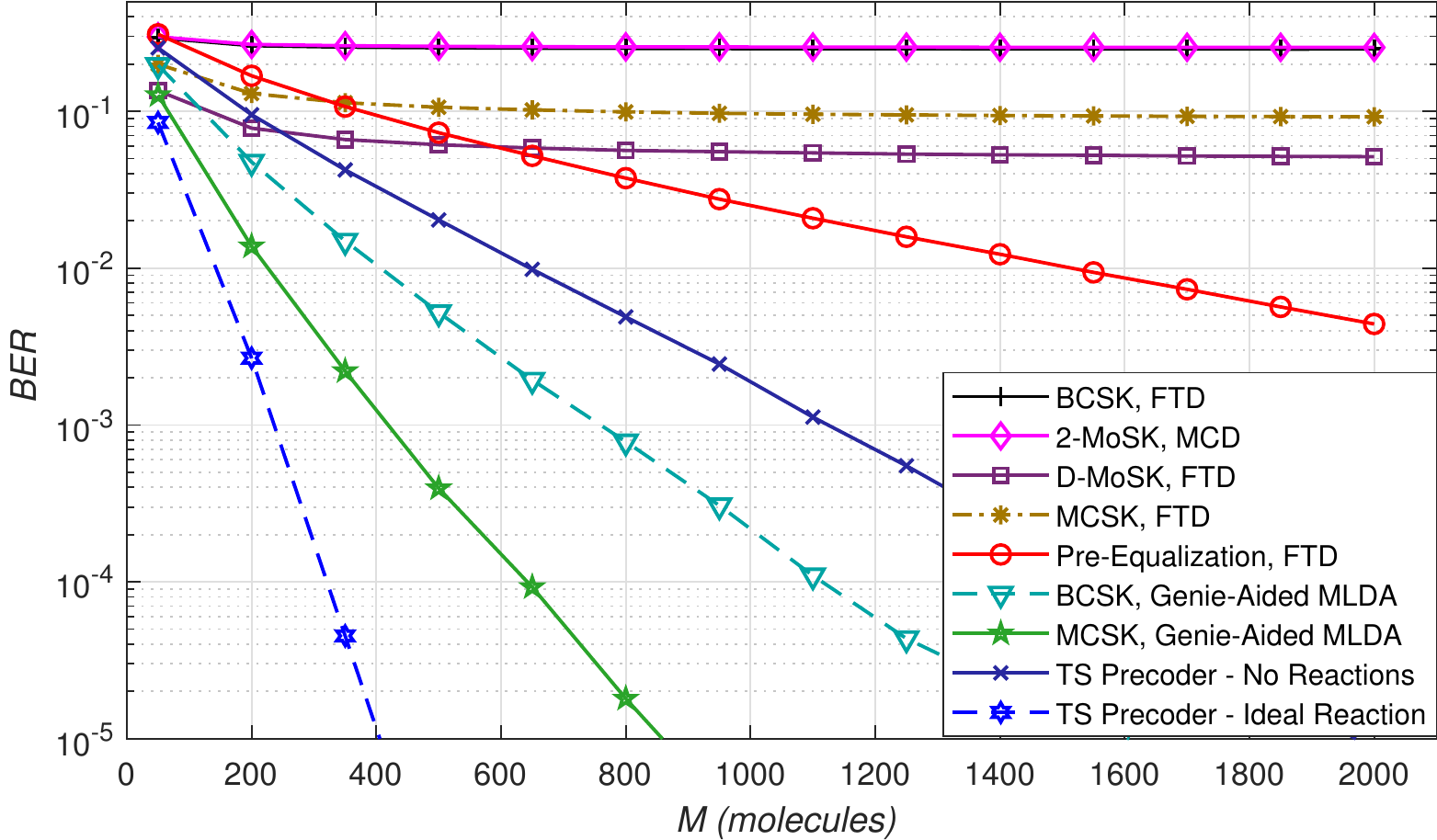} 
	\caption{BER vs. $M$ for various single and dual-molecule schemes. $r_0 = \SI{10}{\micro\meter}$, $r_r = \SI{5}{\micro\meter}$, $D = 80 \frac{\SI{}{\micro\meter\squared}}{\SI{}{\second}}$, data rate $R = 12.5$ bits per second ($t_b = \SI{0.08}{\second}$), $\textrm{SNR} = 15$dB, $L=50$, $S=9\times 10^4$, $N=1$, $\tau = 0$. $\gamma$ values numerically optimized.}
	\label{fig:type_based_sign}
\end{figure}

 As mentioned in Subsection \ref{subsec:type_modulations}, utilizing multiple molecule types arises the question of how to best utilize the available resources. Herein, we seek a numerical answer for this question by comparatively presenting the error performances of select dual-molecule modulation, pre-equalization, and TS-based pre-coding schemes. To serve for this purpose, Figure \ref{fig:type_based_sign} is presented. When generating Figure \ref{fig:type_based_sign}, genie-aided MLDA corresponds to \eqref{eq:MLDA_1} where the ISI in each slot is assumed to be perfectly known at the receiver ($\hat{\mathcal{I}}[k] = \mathcal{I}[k]$). 

\par The results of Figure \ref{fig:type_based_sign} indicates that the pre-coder designed in \cite{typebasedsign_MNK} outperforms other schemes of interest (\textit{i.e.} BCSK, DMoSK \cite{DMoSK_2015}, MCSK \cite{MCSK_arjmandi_2013}, and pre-equalization \cite{aminusb_burcu_2015}). The authors of \cite{typebasedsign_MNK} attribute this improvement to two main reasons: Firstly, comparing the performances of the no-reaction pre-coder and the pre-equalization proposed in \cite{aminusb_burcu_2015}, part of the improvement comes from the signal processing side. Furthermore, the results of Figure \ref{fig:type_based_sign} show that the scheme with the pre-coder and where type-A and type-B molecules react significantly outperforms the no-reaction pre-coding strategy. Therefore, the second improvement comes from the aggressive ISI mitigation provided by the chemical reactions between type-A and type-B molecules. Overall, in addition to using multiple molecule types for modulation or pre-equalization, multi-molecule DiMC allows for an additional degree of freedom in chemical reactions, which we believe can provide promising new applications. 



\section{Concluding Remarks}
\label{sec:conclusion}

\par In this review paper, we have reviewed the state-of-the-art in DiMC transceiver design. Mainly, we have discussed recent information theoretic advancements on DiMC channels and the transceiver signal processing strategies. In addition to the standard detector schemes, we have covered detector schemes that focus on ISI mitigation, acknowledging ISI as the key impediment against high performance DiMC links. Furthermore, we have addressed channel estimation and synchronization in DiMC systems, presenting several non-coherent and/or asynchronous detectors as well. Next, we have talked about the recent advancements in source and channel coding endeavors for DiMC channels, and have provided several design considerations for future research. In the sequel, according to the obtained results and discussed literature, we present several observations that we believe could motivate future studies.
    
\subsection{Use of Chemical Reactions}
\label{subsec:conc_reaction}

\par As in part reflected in this paper, chemical reactions provide performance enhancement to DiMC systems by mitigating ISI (also ILI for molecular MIMO and multi-user interference for multi-access DiMC), signal dependent noise, facilitating physical layer network coding, manipulating the medium for medium-based communications, \textit{etc.}. That said, this unique and truly physical layer facet of multi-molecule DiMC has rarely been addressed in the literature. We believe reaction design is an interesting research direction within DiMC system design, with a very promising performance enhancement potential. Furthermore, it is suitable to be used in conjunction with many existing (and possibly new) signal processing techniques, which can pave the way for various chemical-mathematical joint designs.

\subsection{Pre-processors for Right Tail Suppression}

\par DiMC systems are notorious for their ISI-induced low data rates. In this paper, we presented several transmitter and receiver-side methods that mitigate ISI and improve performance. Unfortunately, powerful ISI mitigation is a challenging task, which implies the difficulty in avoiding the data rate limitations of DiMC. That said, we have exemplified through higher order derivatives that one can \textit{circumvent} the ISI issue by aggressively suppressing the right tail of $f_{hit}(t)$ (see Figure \ref{fig:derivative_fhit}), and provide considerably higher data rates for single-molecule type DiMC systems. Furthermore, the discrete time forward derivative operator is a very simple, non-coherent linear operator which conserves receiver simplicity. Since it can be thought of as a pre-processing block that is placed before the detection stage, it is also compatible for implementation with simple and more complex detector schemes. Overall, the derivative operator's versatility and strong right tail suppression can prove to be very useful in future research towards realizing high data rate DiMC systems. Furthermore, we believe that research towards devising similarly versatile mathematical techniques can provide the tools to push the conventional DiMC data rates considerably faster.

\subsection{Synchronization Error Robustness}
\label{subsec:conc_sync}

\par Especially in nano- to micro-scale applications, synchronization is particularly difficult to achieve due to simple transceiver structures and the nature of the DiMC environments. Many studies focus on encoding release time information within the chemical structure of molecules, which risk being inapplicable for nano- to micro-scale applications. Emphasizing this, we believe devising clock recovery techniques that rely on a single-molecule type is of particular interest going forward. 

\par Furthermore, a large majority of the DiMC detection \& equalization literature assume perfect synchronization and aim for achieving the lowest error probability. However, synchronization in DiMC links is going to be achieved via actual algorithms in practice, which is likely to incur a certain mis-synchronization between the transmitter and the receiver. Emphasizing this imperfection, we believe that robustness against synchronization errors is a major objective for DiMC system design, and transceiver design that specifically targets this objective can be of particular interest going forward.

\subsection{Coding for DiMC}

\par As also mentioned in Section \ref{sec:channel_coding}, coding literature in DiMC is still in its infancy with many open problems. Reciting some of the design considerations listed therein, current challenges, open problems, and design considerations can be briefly listed as follows:
\begin{enumerate}
    \item \textit{New Distance Metrics and Objective Functions:} We had noted that Hamming distance does not fully capture the transition probabilities between codewords. However, direct use of transition probabilities is heavily CSI-dependent (thus weakly generalizable) and computationally complex. Emphasizing these issues, we believe the DiMC coding literature will highly benefit from alternative low-complexity and generalizable cost functions that can be used as proxies to error probability. In fact, in addition to code construction, such cost functions are capable of aiding the researchers in many DiMC equalization, pre-equalization, topological optimization endeavors as well.
    
    \item \textit{Coded-modulation designs:} DiMC coding literature mostly employs BCSK as the modulation scheme and focuses on the code construction aspect of this design consideration. However, the modulation scheme heavily affects the observed ISI and the transmission power of different constellation points (hence the signal dependent noise). Thus, a coupled code-modulation design can exploit the degrees of freedom coming from both aspects (reactions can also be considered in conjunction for multi-molecule DiMC, see Subsection \ref{subsec:conc_reaction}), further enhancing the performance compared to considering them separately.
    
    \item \textit{Robustness against synchronization errors:} In addition, DiMC coding literature also mostly assumes perfect synchronization and a static channel, and aims to achieve the best error performance within this assumption. However, this assumption may not always hold. Motivated by this, we believe robustness against imperfections is these regards can also be a very useful objective to optimize source and channel codes, similar to our discussion in Subsection \ref{subsec:conc_sync}.
\end{enumerate}

\section*{Acknowledgements}

\par This work has been funded in part by one or more of the following grants: ONR N00014-15-1-2550, NSF CCF-1817200, ARO W911NF1910269, Cisco Foundation 1980393, DOE DE-SC0021417, Swedish Research Council 2018-04359, NSF CCF-2008927, and ONR 503400-78050.


\bibliography{elsevier_gursoy_bibtex}

\begin{thebibliography}{100}
\providecommand{\url}[1]{#1}
\csname url@samestyle\endcsname
\providecommand{\newblock}{\relax}
\providecommand{\bibinfo}[2]{#2}
\providecommand{\BIBentrySTDinterwordspacing}{\spaceskip=0pt\relax}
\providecommand{\BIBentryALTinterwordstretchfactor}{4}
\providecommand{\BIBentryALTinterwordspacing}{\spaceskip=\fontdimen2\font plus
\BIBentryALTinterwordstretchfactor\fontdimen3\font minus
  \fontdimen4\font\relax}
\providecommand{\BIBforeignlanguage}[2]{{%
\expandafter\ifx\csname l@#1\endcsname\relax
\typeout{** WARNING: IEEEtran.bst: No hyphenation pattern has been}%
\typeout{** loaded for the language `#1'. Using the pattern for}%
\typeout{** the default language instead.}%
\else
\language=\csname l@#1\endcsname
\fi
#2}}
\providecommand{\BIBdecl}{\relax}
\BIBdecl

\bibitem{firstpaper_lookslike}
T.~Suda, M.~Moore, T.~Nakano, R.~Egashira, A.~Enomoto, S.~Hiyama, and
  Y.~Moritani, ``Exploratory research on molecular communication between
  nanomachines,'' in \emph{Genetic and Evolutionary Computation Conference
  (GECCO), Late Breaking Papers}, vol.~25, 2005, p.~29.

\bibitem{akyildiz_nanocomnet}
I.~F. Akyildiz, F.~Brunetti, and C.~Bl{\'a}zquez, ``Nanonetworks: A new
  communication paradigm,'' \emph{Computer Networks}, vol.~52, no.~12, pp.
  2260--2279, August 2008.

\bibitem{molecularbook}
T.~Nakano, A.~W. Eckford, and T.~Haraguchi, \emph{Molecular
  communication}.\hskip 1em plus 0.5em minus 0.4em\relax Cambridge University
  Press, 2013.

\bibitem{survey_farsad_2016}
N.~{Farsad}, H.~B. {Yilmaz}, A.~{Eckford}, C.~{Chae}, and W.~{Guo}, ``A
  comprehensive survey of recent advancements in molecular communication,''
  \emph{IEEE Communications Surveys Tutorials}, vol.~18, no.~3, pp. 1887--1919,
  third quarter 2016.

\bibitem{survey_schober_2019}
V.~{Jamali}, A.~{Ahmadzadeh}, W.~{Wicke}, A.~{Noel}, and R.~{Schober},
  ``Channel modeling for diffusive molecular communication—a tutorial
  review,'' \emph{Proceedings of the IEEE}, vol. 107, no.~7, pp. 1256--1301,
  July 2019.

\bibitem{oneshot1}
N.~{Farsad}, W.~{Chuang}, A.~{Goldsmith}, C.~{Komninakis}, M.~{Médard},
  C.~{Rose}, L.~{Vandenberghe}, E.~E. {Wesel}, and R.~D. {Wesel}, ``Capacities
  and optimal input distributions for particle-intensity channels,'' \emph{IEEE
  Transactions on Molecular, Biological and Multi-Scale Communications},
  vol.~6, no.~3, pp. 220--232, 2020.

\bibitem{oneshot2}
Y.~Murin, N.~Farsad, M.~Chowdhury, and A.~Goldsmith, ``Exploiting diversity in
  one-shot molecular timing channels via order statistics,'' \emph{IEEE
  Transactions on Molecular, Biological and Multi-Scale Communications},
  vol.~4, no.~1, pp. 14--26, 2018.

\bibitem{abnormality1}
R.~Mosayebi, V.~Jamali, N.~Ghoroghchian, R.~Schober, M.~Nasiri-Kenari, and
  M.~Mehrabi, ``Cooperative abnormality detection via diffusive molecular
  communications,'' \emph{IEEE Transactions on NanoBioscience}, vol.~16, no.~8,
  pp. 828--842, 2017.

\bibitem{abnormality3}
N.~{Varshney}, A.~{Patel}, Y.~{Deng}, W.~{Haselmayr}, P.~K. {Varshney}, and
  A.~{Nallanathan}, ``Abnormality detection inside blood vessels with mobile
  nanomachines,'' \emph{IEEE Transactions on Molecular, Biological and
  Multi-Scale Communications}, vol.~4, no.~3, pp. 189--194, 2018.

\bibitem{localization}
O.~D. {Kose}, M.~C. {Gursoy}, M.~{Saraclar}, A.~E. {Pusane}, and T.~{Tugcu},
  ``Machine learning-based silent entity localization using molecular
  diffusion,'' \emph{IEEE Communications Letters}, vol.~24, no.~4, pp.
  807--810, 2020.

\bibitem{localization2}
S.~Pang and J.~A. Farrell, ``Chemical plume source localization,'' \emph{IEEE
  Transactions on Systems, Man, and Cybernetics, Part B (Cybernetics)},
  vol.~36, no.~5, pp. 1068--1080, 2006.

\bibitem{3Dchar}
H.~B. {Yilmaz}, A.~C. {Heren}, T.~{Tugcu}, and C.~{Chae}, ``Three-dimensional
  channel characteristics for molecular communications with an absorbing
  receiver,'' \emph{IEEE Communications Letters}, vol.~18, no.~6, pp. 929--932,
  2014.

\bibitem{LTI_Poisson}
G.~Aminian, H.~Arjmandi, A.~Gohari, M.~Nasiri-Kenari, and U.~Mitra, ``Capacity
  of diffusion-based molecular communication networks over lti-poisson
  channels,'' \emph{IEEE Transactions on Molecular, Biological and Multi-Scale
  Communications}, vol.~1, no.~2, pp. 188--201, 2015.

\bibitem{arrivalmodel}
H.~B. Yilmaz and C.-B. Chae, ``Arrival modelling for molecular communication
  via diffusion,'' \emph{IET Electronics Letters}, vol.~50, no.~23, pp.
  1667--1669, 2014.

\bibitem{landau1966optimality}
H.~Landau and D.~Slepian, ``On the optimality of the regular simplex code,''
  \emph{Bell System Technical Journal}, vol.~45, no.~8, pp. 1247--1272, 1966.

\bibitem{slepian1976bandwidth}
D.~Slepian, ``On bandwidth,'' \emph{Proceedings of the IEEE}, vol.~64, no.~3,
  pp. 292--300, 1976.

\bibitem{weber2012elements}
C.~L. Weber, \emph{Elements of detection and signal design}.\hskip 1em plus
  0.5em minus 0.4em\relax Springer Science \& Business Media, 2012.

\bibitem{survey_goldsmith_2020}
M.~S. {Kuran}, H.~B. {Yilmaz}, I.~{Demirkol}, N.~{Farsad}, and A.~{Goldsmith},
  ``A survey on modulation techniques in molecular communication via
  diffusion,'' \emph{arXiv preprint arXiv:2011.13056}, 2020.

\bibitem{roder1931amplitude}
H.~Roder, ``Amplitude, phase, and frequency modulation,'' \emph{Proceedings of
  the Institute of Radio Engineers}, vol.~19, no.~12, pp. 2145--2176, 1931.

\bibitem{CSK_MOSK_akyildiz_2011}
M.~S. {Kuran}, H.~B. {Yilmaz}, T.~{Tugcu}, and I.~F. {Akyildiz}, ``Modulation
  techniques for communication via diffusion in nanonetworks,'' in \emph{2011
  IEEE International Conference on Communications (ICC)}, 2011, pp. 1--5.

\bibitem{MCSK_arjmandi_2013}
H.~{Arjmandi}, A.~{Gohari}, M.~N. {Kenari}, and F.~{Bateni}, ``Diffusion-based
  nanonetworking: A new modulation technique and performance analysis,''
  \emph{IEEE Communications Letters}, vol.~17, no.~4, pp. 645--648, April 2013.

\bibitem{FTDadaptivetransmission_mitra_2016}
M.~{Movahednasab}, M.~{Soleimanifar}, A.~{Gohari}, M.~{Nasiri-Kenari}, and
  U.~{Mitra}, ``Adaptive transmission rate with a fixed threshold decoder for
  diffusion-based molecular communication,'' \emph{IEEE Transactions on
  Communications}, vol.~64, no.~1, pp. 236--248, 2016.

\bibitem{ISI_burcu_2015}
B.~Tepekule, A.~E. Pusane, H.~B. Yilmaz, C.-B. Chae, and T.~Tugcu, ``{ISI}
  mitigation techniques in molecular communication,'' \emph{IEEE Transactions
  on Molecular, Biological, and Multi-Scale Communications}, vol.~1, no.~2, pp.
  202--216, June 2015.

\bibitem{DMoSK_2015}
M.~H. Kabir, S.~M.~R. Islam, and K.~S. Kwak, ``{D-MoSK} modulation in molecular
  communications,'' \emph{IEEE Transactions on Nanobioscience}, vol.~14, no.~6,
  pp. 680--683, September 2015.

\bibitem{ISIavoiding_MNK}
H.~Arjmandi, M.~Movahednasab, A.~Gohari, M.~Mirmohseni, M.~Nasiri-Kenari, and
  F.~Fekri, ``{ISI}-avoiding modulation for diffusion-based molecular
  communication,'' \emph{IEEE Transactions on Molecular, Biological and
  Multi-Scale Communications}, vol.~3, no.~1, pp. 48--59, 2016.

\bibitem{isomerMoSK}
N.-R. Kim and C.-B. Chae, ``Novel modulation techniques using isomers as
  messenger molecules for nano communication networks via diffusion,''
  \emph{IEEE Journal on Selected Areas in Communications}, vol.~31, no.~12, pp.
  847--856, 2013.

\bibitem{MaaF_gursoy_2018}
M.~C. Gursoy, A.~E. Pusane, and T.~Tugcu, ``Molecule-as-a-frame: A frame based
  communication approach for nanonetworks,'' \emph{Nano communication
  networks}, vol.~16, pp. 45--59, June 2018.

\bibitem{GMoSK_2020}
X.~{Chen}, Y.~{Huang}, L.~L. {Yang}, and M.~{Wen}, ``Generalized
  molecular-shift keying (gmosk): Principles and performance analysis,''
  \emph{IEEE Transactions on Molecular, Biological and Multi-Scale
  Communications}, vol.~6, no.~3, pp. 168--183, 2020.

\bibitem{runlength_kwak_2014}
S.~Pudasaini, S.~Shin, and K.~S. Kwak, ``Run-length aware hybrid modulation
  scheme for diffusion-based molecular communication,'' in \emph{Proc. IEEE
  Int. Symp. on Commun. Info. Tech. (ISCIT)}, September 2014, pp. 439--442.

\bibitem{enzymes_noel_2014}
A.~{Noel}, K.~C. {Cheung}, and R.~{Schober}, ``Improving receiver performance
  of diffusive molecular communication with enzymes,'' \emph{IEEE Transactions
  on NanoBioscience}, vol.~13, no.~1, pp. 31--43, March 2014.

\bibitem{twowayrelay_MNK}
M.~Farahnak-Ghazani, G.~Aminian, M.~Mirmohseni, A.~Gohari, and
  M.~Nasiri-Kenari, ``On medium chemical reaction in diffusion-based molecular
  communication: A two-way relaying example,'' \emph{IEEE Transactions on
  Communications}, vol.~67, no.~2, pp. 1117--1132, 2018.

\bibitem{aminusb_burcu_2015}
B.~{Tepekule}, A.~E. {Pusane}, M.~S. {Kuran}, and T.~{Tugcu}, ``A novel
  pre-equalization method for molecular communication via diffusion in
  nanonetworks,'' \emph{IEEE Communications Letters}, vol.~19, no.~8, pp.
  1311--1314, August 2015.

\bibitem{typebasedsign_MNK}
R.~Mosayebi, A.~Gohari, M.~Mirmohseni, and M.~Nasiri-Kenari, ``Type-based sign
  modulation and its application for {ISI} mitigation in molecular
  communication,'' \emph{IEEE Transactions on Communications}, vol.~66, no.~1,
  pp. 180--193, 2017.

\bibitem{first_MTC}
A.~W. {Eckford}, ``Nanoscale communication with brownian motion,'' in
  \emph{Annual Conf. Info. Sci. and Syst. (CISS)}, Mar. 2007, pp. 160--165.

\bibitem{AIGN_upperlower}
K.~V. Srinivas, A.~W. Eckford, and R.~S. Adve, ``Molecular communication in
  fluid media: The additive inverse gaussian noise channel,'' \emph{IEEE
  Transactions on Information Theory}, vol.~58, no.~7, pp. 4678--4692, 2012.

\bibitem{MTC_capacitybound}
H.~{Li}, S.~M. {Moser}, and D.~{Guo}, ``Capacity of the memoryless additive
  inverse gaussian noise channel,'' \emph{IEEE J. Sel. Areas Commun.}, vol.~32,
  no.~12, pp. 2315--2329, Dec. 2014.

\bibitem{cooke1947pulse}
D.~Cooke, Z.~Jelonek, A.~Oxford, and E.~Fitch, ``Pulse communication,''
  \emph{Journal of the Institution of Electrical Engineers-Part IIIA:
  Radiocommunication}, vol.~94, no.~11, pp. 83--105, 1947.

\bibitem{scholtz1997impulse}
R.~Scholtz and M.~Z. Win, ``Impulse radio,'' in \emph{Wireless
  Communications}.\hskip 1em plus 0.5em minus 0.4em\relax Springer, 1997, pp.
  245--263.

\bibitem{franz2006generalized}
S.~Franz and U.~Mitra, ``Generalized uwb transmitted reference systems,''
  \emph{IEEE Journal on Selected Areas in Communications}, vol.~24, no.~4, pp.
  780--786, 2006.

\bibitem{franz2007joint}
S.~Franz, C.~Carbonelli, and U.~Mitra, ``Joint semi-blind channel and timing
  estimation for generalized uwb transmitted reference systems,'' \emph{IEEE
  Transactions on Wireless Communications}, vol.~6, no.~1, pp. 180--191, 2007.

\bibitem{murin_PPM_optimal}
Y.~{Murin}, N.~{Farsad}, M.~{Chowdhury}, and A.~{Goldsmith}, ``Optimal
  detection for one-shot transmission over diffusion-based molecular timing
  channels,'' \emph{IEEE Transactions on Molecular, Biological, and Multi-Scale
  Communications}, vol.~4, no.~2, pp. 43--60, 2018.

\bibitem{murin_PPM_otherdetectors}
------, ``Exploiting diversity in one-shot molecular timing channels via order
  statistics,'' \emph{IEEE Transactions on Molecular, Biological, and
  Multi-Scale Communications}, vol.~4, no.~1, pp. 14--26, Mar. 2018.

\bibitem{PPM_original_2011}
N.~Garralda, I.~Llatser, A.~Cabellos-Aparicio, E.~Alarc{\'o}n, and M.~Pierobon,
  ``Diffusion-based physical channel identification in molecular
  nanonetworks,'' \emph{Nano Communication Networks}, vol.~2, no.~4, pp.
  196--204, 2011.

\bibitem{bayram_PPM}
B.~C. Akdeniz, A.~E. Pusane, and T.~Tugcu, ``Position-based modulation in
  molecular communications,'' \emph{Nano communication networks}, vol.~16, pp.
  60--68, 2018.

\bibitem{timing_MNK}
S.~{Aeeneh}, N.~{Zlatanov}, A.~{Gohari}, M.~{Nasiri-Kenari}, and
  M.~{Mirmohseni}, ``Timing modulation for macro-scale molecular
  communication,'' \emph{IEEE Wireless Communications Letters}, vol.~9, no.~9,
  pp. 1356--1360, 2020.

\bibitem{MCPM_mitra_2020}
M.~C. {Gursoy}, D.~{Seo}, and U.~{Mitra}, ``Concentration and position-based
  hybrid modulation scheme for molecular communications,'' in \emph{IEEE
  International Conference on Communications (ICC)}, 2020, pp. 1--6.

\bibitem{index_general}
E.~Basar, M.~Wen, R.~Mesleh, M.~Di~Renzo, Y.~Xiao, and H.~Haas, ``Index
  modulation techniques for next-generation wireless networks,'' \emph{IEEE
  Access}, vol.~5, pp. 16\,693--16\,746, 2017.

\bibitem{flowvelocity_MNK}
M.~Farahnak-Ghazani, M.~Mirmohseni, and M.~Nasiri-Kenari, ``On molecular flow
  velocity meters,'' \emph{to appear in IEEE Transactions on Molecular,
  Biological and Multi-Scale Communications}, 2020.

\bibitem{mediabased_RF}
A.~K. Khandani, ``Media-based modulation: A new approach to wireless
  transmission,'' in \emph{2013 IEEE International Symposium on Information
  Theory}.\hskip 1em plus 0.5em minus 0.4em\relax IEEE, 2013, pp. 3050--3054.

\bibitem{MIMOfirst_akyildiz_2012}
L.~S. Meng, P.~C. Yeh, K.~C. Chen, and I.~F. Akyildiz, ``{MIMO} communications
  based on molecular diffusion,'' in \emph{Proc. IEEE Global Communications
  Conference (GLOBECOM)}, December 2012, pp. 5380--5385.

\bibitem{MIMOtestbed_chae_2016}
B.~H. Koo, C.~Lee, H.~B. Yilmaz, N.~Farsad, A.~Eckford, and C.-B. Chae,
  ``Molecular {MIMO}: From theory to prototype,'' \emph{IEEE Journal on
  Selected Areas in Communication}, vol.~34, no.~3, pp. 600--614, March 2016.

\bibitem{PPM_SMUX}
M.~C. {Gursoy}, E.~{Basar}, A.~E. {Pusane}, and T.~{Tugcu}, ``An {ILI}
  mitigating modulation scheme for molecular {MIMO} communications,'' in
  \emph{International Conference on Telecommunications and Signal Processing
  (TSP)}, 2019, pp. 28--31.

\bibitem{estimationMIMO_spagnolini_2019}
S.~M.~R. {Rouzegar} and U.~{Spagnolini}, ``Diffusive {MIMO} molecular
  communications: Channel estimation, equalization, and detection,'' \emph{IEEE
  Transactions on Communications}, vol.~67, no.~7, pp. 4872--4884, July 2019.

\bibitem{arraygain_damrath_2018}
M.~{Damrath}, H.~B. {Yilmaz}, C.~{Chae}, and P.~A. {Hoeher}, ``Array gain
  analysis in molecular {MIMO} communications,'' \emph{IEEE Access}, vol.~6,
  pp. 61\,091--61\,102, 2018.

\bibitem{alamouti}
S.~M. Alamouti, ``A simple transmit diversity technique for wireless
  communications,'' \emph{IEEE Journal on selected areas in communications},
  vol.~16, no.~8, pp. 1451--1458, 1998.

\bibitem{SM_original}
R.~Y. Mesleh, H.~Haas, S.~Sinanovic, C.~W. Ahn, and S.~Yun, ``Spatial
  modulation,'' \emph{IEEE Transactions on Vehicular Technology}, vol.~57,
  no.~4, pp. 2228--2241, 2008.

\bibitem{index_gursoy_2019}
M.~C. {Gursoy}, E.~{Basar}, A.~E. {Pusane}, and T.~{Tugcu}, ``Index modulation
  for molecular communication via diffusion systems,'' \emph{IEEE Transactions
  on Communications}, vol.~67, no.~5, pp. 3337--3350, May 2019.

\bibitem{spatial_huang_2019}
Y.~{Huang}, M.~{Wen}, L.~L. {Yang}, C.~B. {Chae}, and F.~{Ji}, ``Spatial
  modulation for molecular communication,'' \emph{IEEE Transactions on
  NanoBioscience}, vol.~18, no.~3, pp. 381--395, July 2019.

\bibitem{PPMSM_gursoy_2019}
M.~C. {Gursoy}, E.~{Basar}, A.~E. {Pusane}, and T.~{Tugcu}, ``Pulse
  position-based spatial modulation for molecular communications,'' \emph{IEEE
  Communications Letters}, vol.~23, no.~4, pp. 596--599, April 2019.

\bibitem{misalignments_MIMO}
A.~{Celik}, M.~C. {Gursoy}, E.~{Basar}, A.~E. {Pusane}, and T.~{Tugcu}, ``A
  low-complexity solution to angular misalignments in molecular index
  modulation,'' in \emph{Annual International Symposium on Personal, Indoor and
  Mobile Radio Communications (PIMRC)}, 2019, pp. 1--6.

\bibitem{info_survey}
A.~Gohari, M.~Mirmohseni, and M.~Nasiri-Kenari, ``Information theory of
  molecular communication: Directions and challenges,'' \emph{IEEE Transactions
  on Molecular, Biological and Multi-Scale Communications}, vol.~2, no.~2, pp.
  120--142, 2016.

\bibitem{zeroerror_MNK}
N.~Abadi, A.~Gohari, M.~Mirmohseni, and M.~Nasiri-Kenari, ``On zero-error
  molecular communication with multiple molecule types,'' \emph{IEEE
  Transactions on Communications}, 2020.

\bibitem{time_and_concentration}
F.~Mirkarimi, M.~Mirmohseni, and M.~Nasiri-Kenari, ``On the capacity of the
  joint time and concentration modulation for molecular communications,''
  \emph{arXiv preprint arXiv:2006.13398}, 2020.

\bibitem{wray_timing_1992}
J.~C. Wray, ``An analysis of covert timing channels,'' \emph{Journal of
  Computer Security}, vol.~1, no. 3-4, pp. 219--232, 1992.

\bibitem{moskowitz1992channel}
I.~S. Moskowitz and A.~R. Miller, ``The channel capacity of a certain noisy
  timing channel,'' \emph{IEEE Transactions on Information Theory}, vol.~38,
  no.~4, pp. 1339--1344, 1992.

\bibitem{STC_1994}
------, ``Simple timing channels,'' in \emph{Proc. IEEE Computer Society
  Symposium on Research in Security and Privacy}.\hskip 1em plus 0.5em minus
  0.4em\relax IEEE, 1994, pp. 56--64.

\bibitem{anantharam1996bits}
V.~Anantharam and S.~Verdu, ``Bits through queues,'' \emph{IEEE Transactions on
  Information Theory}, vol.~42, no.~1, pp. 4--18, 1996.

\bibitem{AIGN}
K.~V. {Srinivas}, A.~W. {Eckford}, and R.~S. {Adve}, ``Molecular communication
  in fluid media: The additive inverse gaussian noise channel,'' \emph{IEEE
  Transactions on Information Theory}, vol.~58, no.~7, pp. 4678--4692, 2012.

\bibitem{diffusion_based_MTC}
N.~Farsad, Y.~Murin, A.~Eckford, and A.~Goldsmith, ``On the capacity of
  diffusion-based molecular timing channels,'' in \emph{IEEE International
  Symposium on Information Theory (ISIT)}.\hskip 1em plus 0.5em minus
  0.4em\relax IEEE, 2016, pp. 1023--1027.

\bibitem{inscribed_matter}
C.~Rose and I.~S. Mian, ``Inscribed matter communication: Part i,'' \emph{IEEE
  Transactions on Molecular, Biological and Multi-Scale Communications},
  vol.~2, no.~2, pp. 209--227, 2016.

\bibitem{MTC_finitelifetime}
N.~{Farsad}, Y.~{Murin}, A.~W. {Eckford}, and A.~{Goldsmith}, ``Capacity limits
  of diffusion-based molecular timing channels with finite particle lifetime,''
  \emph{IEEE Transactions on Molecular, Biological and Multi-Scale
  Communications}, vol.~4, no.~2, pp. 88--106, 2018.

\bibitem{lapidoth_onthe}
A.~{Lapidoth} and S.~M. {Moser}, ``On the capacity of the discrete-time poisson
  channel,'' \emph{IEEE Transactions on Information Theory}, vol.~55, no.~1,
  pp. 303--322, 2009.

\bibitem{lapidoth_lowinput}
A.~{Lapidoth}, J.~H. {Shapiro}, V.~{Venkatesan}, and L.~{Wang}, ``The
  discrete-time poisson channel at low input powers,'' \emph{IEEE Transactions
  on Information Theory}, vol.~57, no.~6, pp. 3260--3272, 2011.

\bibitem{bacterialcable_capacity}
N.~Michelusi and U.~Mitra, ``Capacity of electron-based communication over
  bacterial cables: the full-csi case,'' \emph{IEEE Transactions on Molecular,
  Biological and Multi-Scale Communications}, vol.~1, no.~1, pp. 62--75, 2015.

\bibitem{chen2005capacity}
J.~Chen and T.~Berger, ``The capacity of finite-state markov channels with
  feedback,'' \emph{IEEE Transactions on Information Theory}, vol.~51, no.~3,
  pp. 780--798, 2005.

\bibitem{michelusi2016queuing}
N.~Michelusi, J.~Boedicker, M.~El-Naggar, and U.~Mitra, ``Queuing models for
  abstracting interactions in bacterial communities,'' \emph{IEEE Journal on
  Selected Areas in Communications}, vol.~34, no.~3, pp. 584--599, 2016.

\bibitem{michelusi2014stochastic}
N.~Michelusi, S.~Pirbadian, M.~El-Naggar, and U.~Mitra, ``A stochastic model
  for electron transfer in bacterial cables,'' \emph{IEEE Journal on Selected
  Areas in Communications}, vol.~32, no.~12, pp. 2402--2416, 2014.

\bibitem{viterbi1967error}
A.~Viterbi, ``Error bounds for convolutional codes and an asymptotically
  optimum decoding algorithm,'' \emph{IEEE Transactions on Information Theory},
  vol.~13, no.~2, pp. 260--269, 1967.

\bibitem{adaptivethreshold_damrath_2016}
M.~Damrath and P.~A. Hoeher, ``Low-complexity adaptive threshold detection for
  molecular communication,'' \emph{IEEE Transactions on NanoBioscience},
  vol.~15, no.~3, pp. 200--208, 2016.

\bibitem{memorysamplingrate_mitra_2014}
R.~{Mosayebi}, H.~{Arjmandi}, A.~{Gohari}, M.~{Nasiri-Kenari}, and U.~{Mitra},
  ``Receivers for diffusion-based molecular communication: Exploiting memory
  and sampling rate,'' \emph{IEEE Journal on Selected Areas in Communications},
  vol.~32, no.~12, pp. 2368--2380, 2014.

\bibitem{MLSD_DFE_akan}
D.~Kilinc and O.~B. Akan, ``Receiver design for molecular communication,''
  \emph{IEEE Journal on Selected Areas in Communications}, vol.~31, no.~12, pp.
  705--714, December 2013.

\bibitem{wald_SPRT}
A.~Wald, ``Sequential tests of statistical hypotheses,'' \emph{The annals of
  mathematical statistics}, vol.~16, no.~2, pp. 117--186, 1945.

\bibitem{sprt_mitra}
T.~{Tung} and U.~{Mitra}, ``Synchronization error robust transceivers for
  molecular communication,'' \emph{IEEE Transactions on Molecular, Biological
  and Multi-Scale Communications}, vol.~5, no.~3, pp. 207--221, 2019.

\bibitem{lin_derivative_2018}
H.~{Yan}, G.~{Chang}, Z.~{Ma}, and L.~{Lin}, ``Derivative-based signal
  detection for high data rate molecular communication system,'' \emph{IEEE
  Commun. Lett.}, vol.~22, no.~9, pp. 1782--1785, Sep. 2018.

\bibitem{gursoymitra_derivative_2020}
M.~C. {Gursoy} and U.~{Mitra}, ``Higher order derivatives: Improved
  pre-processing and receivers for molecular communications,'' in \emph{Proc.
  IEEE Global Communications Conference (GLOBECOM)}, December 2020.

\bibitem{bandedMLSD}
A.~Kavcic and J.~M. Moura, ``The viterbi algorithm and markov noise memory,''
  \emph{IEEE Transactions on information theory}, vol.~46, no.~1, pp. 291--301,
  2000.

\bibitem{survey_akan_2019}
M.~{Kuscu}, E.~{Dinc}, B.~A. {Bilgin}, H.~{Ramezani}, and O.~B. {Akan},
  ``Transmitter and receiver architectures for molecular communications: A
  survey on physical design with modulation, coding, and detection
  techniques,'' \emph{Proceedings of the IEEE}, vol. 107, no.~7, pp.
  1302--1341, July 2019.

\bibitem{estimation_noel_2015}
A.~Noel, K.~C. Cheung, and R.~Schober, ``Joint channel parameter estimation via
  diffusive molecular communication,'' \emph{IEEE Transactions on Molecular,
  Biological and Multi-Scale Communications}, vol.~1, no.~1, pp. 4--17, 2015.

\bibitem{estimation_jamali_2016}
V.~{Jamali}, A.~{Ahmadzadeh}, C.~{Jardin}, H.~{Sticht}, and R.~{Schober},
  ``Channel estimation for diffusive molecular communications,'' \emph{IEEE
  Transactions on Communications}, vol.~64, no.~10, pp. 4238--4252, October
  2016.

\bibitem{semiblind_estimation_2020}
S.~Abdallah and A.~M. Darya, ``Semi-blind channel estimation for diffusive
  molecular communication,'' \emph{IEEE Communications Letters}, vol.~24,
  no.~11, pp. 2503--2507, 2020.

\bibitem{ANNdetection}
N.~Farsad and A.~Goldsmith, ``Neural network detection of data sequences in
  communication systems,'' \emph{IEEE Transactions on Signal Processing},
  vol.~66, no.~21, pp. 5663--5678, 2018.

\bibitem{farsad_testbed}
N.~Farsad, W.~Guo, and A.~W. Eckford, ``Tabletop molecular communication: Text
  messages through chemical signals,'' \emph{PloS one}, vol.~8, no.~12, p.
  e82935, 2013.

\bibitem{drift_lin_2017}
L.~{Lin}, J.~{Zhang}, M.~{Ma}, and H.~{Yan}, ``Time synchronization for
  molecular communication with drift,'' \emph{IEEE Communications Letters},
  vol.~21, no.~3, pp. 476--479, 2017.

\bibitem{AIGN_lin_2015}
L.~{Lin}, C.~{Yang}, M.~{Ma}, and S.~{Ma}, ``Diffusion-based clock
  synchronization for molecular communication under inverse gaussian
  distribution,'' \emph{IEEE Sensors Journal}, vol.~15, no.~9, pp. 4866--4874,
  2015.

\bibitem{a_clock_linlin}
L.~Lin, C.~Yang, M.~Ma, S.~Ma, and H.~Yan, ``A clock synchronization method for
  molecular nanomachines in bionanosensor networks,'' \emph{IEEE Sensors
  Journal}, vol.~16, no.~19, pp. 7194--7203, 2016.

\bibitem{SIMO_sync}
Z.~{Luo}, L.~{Lin}, W.~{Guo}, S.~{Wang}, F.~{Liu}, and H.~{Yan}, ``One symbol
  blind synchronization in simo molecular communication systems,'' \emph{IEEE
  Wireless Communications Letters}, vol.~7, no.~4, pp. 530--533, 2018.

\bibitem{TNB_schober_2017}
V.~{Jamali}, A.~{Ahmadzadeh}, and R.~{Schober}, ``Symbol synchronization for
  diffusion-based molecular communications,'' \emph{IEEE Transactions on
  NanoBioscience}, vol.~16, no.~8, pp. 873--887, 2017.

\bibitem{sync_twomolecules_2019}
M.~{Mukherjee}, H.~B. {Yilmaz}, B.~B. {Bhowmik}, J.~{Lloret}, and Y.~{Lv},
  ``Synchronization for diffusion-based molecular communication systems via
  faster molecules,'' in \emph{ICC 2019 - 2019 IEEE International Conference on
  Communications (ICC)}, 2019, pp. 1--5.

\bibitem{blind_2013}
H.~{ShahMohammadian}, G.~G. {Messier}, and S.~{Magierowski}, ``Blind
  synchronization in diffusion-based molecular communication channels,''
  \emph{IEEE Communications Letters}, vol.~17, no.~11, pp. 2156--2159, 2013.

\bibitem{mobile_sync}
L.~{Huang}, L.~{Lin}, F.~{Liu}, and H.~{Yan}, ``Clock synchronization for
  mobile molecular communication systems,'' \emph{IEEE Transactions on
  NanoBioscience}, pp. 1--1, 2020.

\bibitem{training_pcyeh}
B.-K. Hsu, P.-C. Chou, C.-H. Lee, and P.-C. Yeh, ``Training-based
  synchronization for quantity-based modulation in inverse gaussian channels,''
  in \emph{2017 IEEE International Conference on Communications (ICC)}.\hskip
  1em plus 0.5em minus 0.4em\relax IEEE, 2017, pp. 1--6.

\bibitem{async_peak_noel}
A.~Noel and A.~W. Eckford, ``Asynchronous peak detection for demodulation in
  molecular communication,'' in \emph{2017 IEEE International Conference on
  Communications (ICC)}.\hskip 1em plus 0.5em minus 0.4em\relax IEEE, 2017, pp.
  1--6.

\bibitem{bayram_sync}
B.~C. {Akdeniz}, A.~E. {Pusane}, and T.~{Tugcu}, ``Optimal reception delay in
  diffusion-based molecular communication,'' \emph{IEEE Communications
  Letters}, vol.~22, no.~1, pp. 57--60, 2018.

\bibitem{hamming_leeson_2012}
M.~S. Leeson and M.~D. Higgins, ``Forward error correction for molecular
  communications,'' \emph{Nano Communication Networks}, vol.~3, no.~3, pp.
  161--167, 2012.

\bibitem{convolutional1}
M.~U. Mahfuz, D.~Makrakis, and H.~T. Mouftah, ``Performance analysis of
  convolutional coding techniques in diffusion-based concentration-encoded pam
  molecular communication systems,'' \emph{BioNanoScience}, vol.~3, no.~3, pp.
  270--284, 2013.

\bibitem{ISIfree_2013}
P.-J. Shih, C.-H. Lee, P.-C. Yeh, and K.-C. Chen, ``Channel codes for
  reliability enhancement in molecular communication,'' \emph{IEEE Journal on
  Selected Areas in Communications}, vol.~31, no.~12, pp. 857--867, 2013.

\bibitem{SOCC_2015}
Y.~Lu, M.~D. Higgins, and M.~S. Leeson, ``Self-orthogonal convolutional codes
  ({SOCC}s) for diffusion-based molecular communication systems,'' in
  \emph{2015 IEEE International Conference on Communications (ICC)}.\hskip 1em
  plus 0.5em minus 0.4em\relax IEEE, 2015, pp. 1049--1053.

\bibitem{RScodes_mitra_2019}
M.~B. {Dissanayake}, Y.~{Deng}, A.~{Nallanathan}, M.~{Elkashlan}, and
  U.~{Mitra}, ``Interference mitigation in large-scale multiuser molecular
  communication,'' \emph{IEEE Transactions on Communications}, vol.~67, no.~6,
  pp. 4088--4103, 2019.

\bibitem{kislal_2020}
A.~O. Kislal, B.~C. Akdeniz, C.~Lee, A.~E. Pusane, T.~Tugcu, and C.-B. Chae,
  ``{ISI}-mitigating channel codes for molecular communication via diffusion,''
  \emph{IEEE Access}, vol.~8, pp. 24\,588--24\,599, 2020.

\bibitem{MEC_2014}
C.~Bai, M.~S. Leeson, and M.~D. Higgins, ``Minimum energy channel codes for
  molecular communications,'' \emph{Electronics Letters}, vol.~50, no.~23, pp.
  1669--1671, 2014.

\bibitem{moco_distance_2012}
P.-Y. Ko, Y.-C. Lee, P.-C. Yeh, C.-h. Lee, and K.-C. Chen, ``A new paradigm for
  channel coding in diffusion-based molecular communications: Molecular coding
  distance function,'' in \emph{IEEE Global Communications Conference
  (GLOBECOM)}.\hskip 1em plus 0.5em minus 0.4em\relax IEEE, 2012, pp.
  3748--3753.

\bibitem{ISIaware_oguz_2019}
A.~O. Kislal, H.~B. Yilmaz, A.~E. Pusane, and T.~Tugcu, ``{ISI}-aware channel
  code design for molecular communication via diffusion,'' \emph{IEEE
  Transactions on NanoBioscience}, vol.~18, no.~2, pp. 205--213, 2019.

\end{thebibliography}

\end{document}